\documentclass[a4paper,9pt,numreference,epsfig,cite]{article}

\usepackage[latin1]{inputenc}
\usepackage{textcomp}
\usepackage{graphicx}
\usepackage{xcolor}

\newcommand{\stt}{\small\tt}

\newcommand{\Abstract}[1]{\begin{center}{\stt ABSTRACT}\end{center}{#1}
\vskip 0.4in}

\def\Journal#1&#2&#3(#4){\unskip, #1~{\bf #2} (#4) #3}

\def\U4S{\Upsilon (\mbox{4S})}

\def\b0b0{${\rm B^{o}\overline{B^o}}$}

\newcommand{\be}{\begin{equation}}
\newcommand{\ee}{\end{equation}}
\newcommand{\ba}{\begin{array}{c}}
\newcommand{\ea}{\end{array}}
\newcommand{\beqn}{\begin{eqnarray}}
\newcommand{\eeqn}{\end{eqnarray}}

\title{Study of  CP violation in $B^\pm$ decays to $\overline{D^0}(D^0) K^\pm$ at FCCee\protect\\}

\author{R. Aleksan$^1$, L. Oliver$^2$ and E. Perez$^3$ \\
\footnotesize $^1$ IRFU, CEA, Universit\'e Paris-Saclay, 91191 Gif-sur-Yvette cedex, France \\
\footnotesize $^2$ IJCLab, P\^ole Th\'eorie, CNRS/IN2P3 et Universit\'e Paris-Saclay, b\^at. 210, 91405 Orsay, France \\
\footnotesize $^3$ CERN, EP Department, Geneva, Switzerland} 

\begin{document}

\maketitle

\Abstract{
The great progress made recently in the sector of Flavor Physics has enabled to establish CP violation in the B-meson decays. The unitarity triangle derived from the unitarity relation $V_{ub}^* V_{ud}  + V_{cb}^* V_{cd}  + V_{tb}^* V_{td} = 0$ has been measured very precisely. To further asses our understanding of CP violation, it would be useful to carry out similar measurement of other triangles. In this note, we investigate the triangle derived from the relation $V_{ub}^* V_{us}  + V_{cb}^* V_{cs}  + V_{tb}^* V_{ts} = 0$.
Two angles of this triangle ($\alpha_s$ and $\beta_s$) could be measured very accurately at FCCee using the decays $B_s(\overline{B_s})\rightarrow D^\pm_sK^\mp$ and $B_s(\overline{B_s})\rightarrow J/\psi \phi$ respectively, as discussed elsewhere by us.
This note concentrates on the measurement of the third angle $\gamma_s$ using the modes $B^\pm \to \overline{D^0}(D^0)K^\pm$. We show that a direct measurement of the angle $\gamma_s$ is possible with some specific $B^\pm$ decays with an estimated resolution of the order of 1$^\circ$.}

 
\section{Introduction}
    
\noindent The purpose of this note is to discuss the general features of the reactions $B^+ \to \overline{D^0}(D^0)K^+$, in particular with respect to the interference effects which one expects in the Standard Model (SM).\par

\vskip 3 truemm

Should there be only 3 families, the CKM matrix elements obey the following triangular unitarity relations,
$$\qquad \qquad \qquad UT_{bd} \equiv V_{ub}^* V_{ud} + V_{cb}^* V_{cd} + V_{tb}^* V_{td} = 0 \qquad \qquad \qquad \ \ \  $$
\be
\qquad \qquad UT_{sb} \equiv V_{ub}^* V_{us} + V_{cb}^* V_{cs} + V_{tb}^* V_{ts} = 0 \ \  \qquad \qquad
\label{eq:1}
\ee 
$$\qquad \qquad \qquad UT_{ds} \equiv V_{us}^* V_{ud} + V_{cs}^* V_{cd} + V_{ts}^* V_{td} = 0 \qquad \qquad \qquad \ \ \  $$

\vskip 3 truemm

In the first equation (\ref{eq:1}), $UT_{bd}$ is known as the Unitarity Triangle, with the three sides of the same order. The second equation corresponds to a significantly flatter triangle, while the third one is almost completely flat. These triangles are visualized in Fig. 1.\par 

\begin{center}

\includegraphics[scale=1.3]{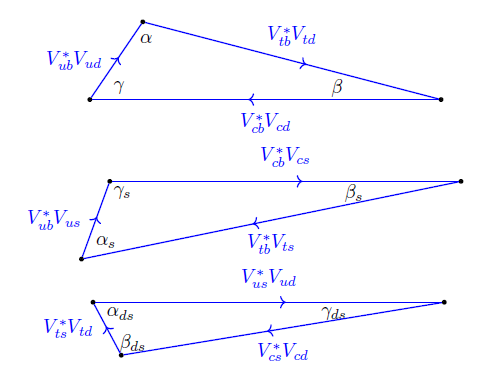}

\vskip 3 truemm

Fig. 1. Unitarity Triangle UT$_{bd}$ involving the $3^{rd}$ and $1^{st}$ columns (top), Unitarity Triangle UT$_{sb}$ involving the $2^{nd}$ and $3^{rd}$ columns (center) and Unitarity Triangle UT$_{ds}$ involving the $1^{nd}$ and $2^{nd}$ columns (bottom) of the CKM matrix (not to scale).

\end{center}

\vskip 3 truemm

There are also three additional triangles, but they are very similar to those above. In the SM, the CKM matrix has only four independent parameters. Therefore the angles of these triangles can be expressed in terms of four angles. The first relation in (\ref{eq:1}), with the three sides of the same order, has been studied extensively. However the other ones would deserve to be studied in detail as well, in order to investigate further the consistency of the SM.\par
We define the angles of these triangles as 
$$\alpha =  \arg \left(-\frac{  V_{tb}^* V_{td} }{ V_{ub}^* V_{ud} }\right) \ , \ \ \ \beta = \arg \left(-{ V_{cb}^* V_{cd} \over V_{tb}^* V_{td} }\right)  \ , \ \ \ \gamma = \arg \left(-{  V_{ub}^* V_{ud} \over V_{cb}^* V_{cd} }\right) \qquad$$
\be
\alpha_s =  \arg \left(-\frac{  V_{ub}^* V_{us} }{ V_{tb}^* V_{ts} }\right) \ , \ \ \ \beta_s =   \arg \left(-{ V_{tb}^* V_{ts} \over V_{cb}^* V_{cs} }\right) \ , \ \ \ \gamma_s =  \arg \left(-{  V_{cb}^* V_{cs} \over V_{ub}^* V_{us} }\right)
\label{eq:2}
\ee
$$\alpha_{ds} = \arg \left(-\frac{  V_{us}^* V_{ud} }{ V_{ts}^* V_{td} }\right) \ , \ \ \ \beta_{ds} =  \arg \left(-{  V_{ts}^* V_{td} \over V_{cs}^* V_{cd} }\right) \ , \ \ \ \gamma_{ds} =   \arg \left(-{ V_{cs}^* V_{cd} \over V_{us}^* V_{ud} }\right) \qquad$$

\vskip 3 truemm

After having studied the direct determination of the angles $\alpha_s$ and $\beta_s$ at FCCee in ref. \cite{ALEKSAN}, in the present paper we are interested and we focus on the direct determination of the angle $\gamma_s$ of the "rather flat" triangle $UT_{sb}$.  By {\it direct determination} of one angle we mean measurement of the angle without making use of its relations to other angles of the triangles in the SM \cite{AKL}.

\vskip 10pt

\section{Study of interference effect for the reactions $B^+ \to \overline{D^0}(D^0)K^+$ and their CP conjugates.}

\noindent The decays $B^+ \to \overline{D^0}K^+$ and $B^+ \to D^0K^+$, as well as their CP conjugates, can interfere in some specific cases. These decays are penguin-free and their interference may lead to clean measurement of the CKM phase involved and will be discussed in the following.

\subsection{Direct CP violation in $B^+$ decays}

\noindent We first investigate how CP violation can be generated with the modes $B^+ \to \overline{D^0}(D^0)K^+$. The decay diagrams for these mode are  shown in Fig. 2 and Fig. 3.

\vskip 4truemm

\includegraphics[scale=1.1]{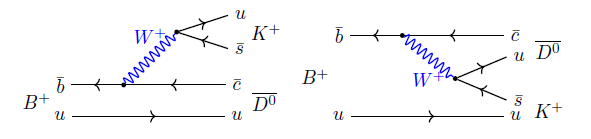}

Fig. 2. The leading Feynman diagrams for the $B^+$ decay for the final state $\overline{D^0}K^+$. The product of CKM elements involved is $V_{cb}^*V_{us}$.

\vskip 4truemm

\includegraphics[scale=1.1]{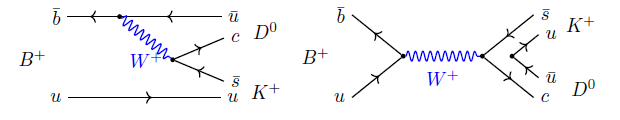}

Fig. 3. The leading Feynman diagrams for the $B^+$ decay for the final state $D^0K^+$. The product of CKM elements involved is $V_{ub}^*V_{cs}$.

\vskip 4truemm

\noindent To generate CP violation in the Standard Model (SM), at least two interfering amplitudes with different strong phases and different weak phases are required.

\noindent The diagrams in Fig. 2 and in Fig. 3 have distinct CKM weak phases and different strong phases and thus their interference would generate CP violation. This is not always possible since, in general, the final states of $\overline{D^0}$ and $D^0$ are different, however there are two ways to get around this by selecting specific $\overline{D^0}(D^0)$ decay final states. 
\begin{enumerate}
\item Using Doubly Cabibbo Suppressed (DCS) final states for $\overline{D^0}$ decays, e.g. $\overline{D^0}\rightarrow K^-\pi^+$, while using the Cabibbo Allowed (CA) final states for $D^0$ decays, i.e.  $D^0\rightarrow K^-\pi^+$. Obviously, similarly, one may use CA states for $\overline{D^0}$ decays and DCS states for $D^0$ decays.
\item Using  $\overline{D^0}(D^0)$ decays to CP eigenstates, e.g. $\overline{D^0}(D^0)\rightarrow \pi^+\pi^-$, $K^+K^-$ or $K_s\pi^0$.
\end{enumerate}
\noindent One can easily increase the interesting final states by adding $\pi^0$ in the final states.

\noindent Let us write for illustration two generic interfering amplitudes and their CP conjugates, following the diagrams of Figs. 2 and 3.
$$A_1 = A(B^+ \to \overline{D}^0 K^+ \to f_i K^+) = a_1e^{i\alpha_1} e^{i\Phi_1}$$
$$A_2 = A(B^+ \to D^0 K^+ \to f_i K^+) = a_2e^{i\alpha_2} e^{i\Phi_2}$$

$$\overline{A_1} = A(B^- \to D^0 K^- \to \overline{f}_i K^-) = a_1e^{i\alpha_1} e^{-i\Phi_1}$$
\be
\qquad \qquad \qquad\overline{A_2} = A(B^- \to \overline{D}^0 K^- \to \overline{f}_i K^-) = a_2e^{i\alpha_2} e^{-i\Phi_2} \ \qquad \qquad
\label{eq:3}
\ee

\noindent where $a_1(a_2)$ are real numbers, $\alpha_1(\alpha_2)$ are the strong phases and $\Phi_1(\Phi_2)$ are the weak phases. 
With the notation $\Delta =\alpha_1 -\alpha_2$ and  $\Phi =\Phi_1-\Phi_2$, one can write the partial widths 

$$|A|^2 = |A_1 +A_2|^2 =  a_1^2 + a_2^2 +2a_1a_2 [\cos\Delta \cos\Phi - \sin\Delta \sin\Phi] \qquad \qquad \ \ \ $$ 
\be  
|\overline{A}|^2 = |\overline{A_1} +\overline{A_2}|^2 = a_1^2 + a_2^2 +2a_1a_2 [\cos\Delta \cos\Phi + \sin\Delta \sin\Phi] \qquad \qquad
\label{eq:4}
\ee

\vskip 3truemm

\noindent Should one know $a_1^2$ and $a_2^2$, one is left with two equations with two unknowns, $\Delta$ and $\Phi$.
However this is not trivial as discussed below.\par 

\vskip 3truemm

Let us rewrite equations (4) for the final state corresponding to diagrams in Fig. 2 and Fig. 3, with $D^0$ and $\overline{D}^0$ decaying into a common final state $f_i$.\par 
The amplitudes $A_1$ and $A_2$ will respectively correspond e.g. to the decay chains $B^+ \to \overline{D}^0 K^+ \to f_i K^+$ and $B^+ \to D^0 K^+ \to f_i K^+$. We will have then, factorizing $\Gamma[B^+\to \overline{D^0}K^+ \to f_i K^+]$ for the decay $\Gamma\left(B^+ \to (f_i)_D K^+\right)$, and proceeding likewise for its CP conjugate :
$$\mid A\mid^2\ \equiv \Gamma\left(B^+ \to (f_i)_D K^+\right) = \Gamma[B^+\to \overline{D^0}K^+  \to f_i K^+] \qquad \qquad \qquad$$  
$$\times [1 + {\cal R}_{B\to f_iK}^2 +2{\cal R}_{B\to f_iK} (\cos\Delta_{f_i} \cos\Phi_{f_i} - \sin\Delta_{f_i} \sin\Phi_{f_i} )]$$
$$\mid \overline{A}\mid^2\ \equiv \Gamma\left(B^- \to (\overline{f}_i)_D K^-\right) = \Gamma[B^-\to D^0K^-  \to \overline{f_i}K^-] \qquad \qquad \qquad $$
\be 
\ \ \ \times [1 + \overline{{\cal R}}_{B\to \overline{f_iK}}^2 +2\overline{{\cal R}}_{B\to \overline{f_iK}}(\cos\Delta_{\overline{f_i}} \cos\Phi_{\overline{f_i}} - \sin\Delta_{\overline{f_i}} \sin\Phi_{\overline{f_i}} )]\qquad \qquad 
\label{eq:5}
\ee

\vskip 2truemm

\noindent where $f_i(\overline{f_i})$ is a final state {\it common} to $\overline{D^0}$ and $D^0$, the strong and weak phases satisfy
\be
\qquad \ \ \ \Delta_{f_i} = \Delta_{\overline{f_i}} =\alpha_1 -\alpha_2 \ , \qquad \Phi_{f_i} = -\Phi_{\overline{f_i}} =\Phi_1 -\Phi_2 \qquad \qquad \qquad 
\label{eq:6} 
\ee

\vskip 2truemm

\noindent and the ratios in  (\ref{eq:5}) are defined by

$${\cal R}_{B\to f_iK} = \sqrt {{\Gamma[B^+\to D^0K^+  \to f_i K^+] \over \Gamma[B^+\to \overline{D^0}K^+  \to f_i K^+]}}  \qquad \qquad \qquad$$
$$= \sqrt {{\Gamma[B^+\to D^0K^+] \over \Gamma[B^+\to \overline{D^0}K^+]}} \times 
\sqrt {{\Gamma[D^0  \to f_i] \over \Gamma[\overline{D^0}  \to f_i]}} \equiv {\cal R} \times {\cal R}_{D\to f_i}  \qquad \qquad \qquad $$

\vskip 1truemm

$$\overline{{\cal R}}_{B\to \overline{f_iK}} = \sqrt {{\Gamma[B^-\to \overline{D^0}K^-  \to \overline{f_i}K^-] \over \Gamma[B^-\to D^0K^-  \to \overline{f_i}K^-]}} \qquad \qquad \qquad \qquad $$
\be
\qquad = \sqrt {{\Gamma[B^-\to \overline{D^0}K^-] \over \Gamma[B^-\to D^0K^-]}} \times 
\sqrt {{\Gamma[\overline{D^0}  \to \overline{f_i}] \over \Gamma[D^0  \to \overline{f_i]}}} \equiv \overline{{\cal R}} \times \overline{{\cal R}}_{D\to \overline{f_i}} \qquad \qquad \qquad \ \ \ \ 
\label{eq:7}
\ee

\vskip 2truemm

\noindent One has the relations
\be
\qquad \qquad {\cal R} = \overline{{\cal R}} \ , \qquad \qquad {\cal R}_{D\to f_i} = \overline{{\cal R}}_{D\to \overline{f_i}} = {1\over {\cal R}_{D\to \overline{f_i}}} \qquad \qquad \qquad \ \ \ 
\label{eq:8}
\ee

\noindent that imply
\be
\qquad \qquad \qquad\qquad \qquad {\cal R}_{B\to f_iK} = \overline{{\cal R}}_{B\to \overline{f_iK}}\qquad \qquad \qquad \qquad \qquad \qquad 
\label{eq:9}
\ee

\noindent and from eqns. (\ref{eq:5})-(\ref{eq:9}) one gets finally, after some algebra,
$$ \qquad \qquad \Gamma\left(B^- \to (\overline{f}_i)_D K^-\right) = \Gamma[B^+\to \overline{D^0}K^+  \to f_iK^+] \qquad \qquad \qquad \qquad \qquad \qquad \qquad$$  
\be
\qquad \times [1 + {\cal R}_{B\to f_iK}^2 +2{\cal R}_{B\to f_iK}(\cos\Delta_{f_i} \cos\Phi_{f_i} + \sin\Delta_{f_i} \sin\Phi_{f_i} )] \qquad \qquad 
\label{eq:10}
\ee

From an experimental point of view, $|A|^2\ (|\overline{A}|^2)$ is the number of events 
\noindent $B^+ \to (f_i)_D K^+\ (B^- \to (\overline{f_i})_D K^-)$ (i.e. the measured signal), while $\Gamma[B^+\to \overline{D^0}K^+ \to f_i K^+]$ is the number of $B^+\to \overline{D^0}K^+$ followed by $\overline{D^0}\to f_i$ decays. 

\vskip 10pt
Looking at the first eqn. (\ref{eq:5}) and eqn. (\ref{eq:10}), we have 4 unknows, $\Gamma[B^+\to \overline{D^0}K^+ \to f_i K^+]$, ${\cal R}_{B\to f_iK}$, $\Delta_{f_i}$ and $\Phi_{f_i}$.\par

\vskip 3 truemm
 
With presently operating and future accelerators like LHC, superKEKB and FCC, it should be possible to measure ${\cal R}$ and  ${\cal R}_{D\to f_i}$, and thus ${\cal R}_{B\to f_iK}$, with high accuracy for most of the relevant final states.\par 

\vskip 3 truemm

Indeed, one has,
$$\qquad {\cal R}^2 = {Br(B^+ \to D^0K^+) \over Br(B^+ \to \overline{D^0}K^+)} =
 {Br(B^- \to \overline{D^0}K^-) \over Br(B^- \to D^0K^-)} \qquad \qquad \qquad \qquad$$
\be
\qquad\qquad  {\cal R}_{D\to f_i}^2 = { Br(D^0  \to f_i) \over  Br(\overline{D^0} \to f_i) } = 
 {Br(\overline{D^0}  \to \overline {f_i}) \over  Br(D^0 \to \overline{f_i})} \qquad \qquad \qquad \qquad \qquad
\label{eq:11}
\ee

\noindent and these fractions can be determined experimentally, as we argue below\par 

\vskip 3 truemm

- ${\cal R}$ is measured by using semileptonics decays of the $\overline{D^0}(D^0)$. Indeed, the decays $D^0\rightarrow \ell^+\nu K^-$ and $\overline{D^0}\rightarrow \ell^-\bar{\nu}K^+$, the Branching Fractions of which are known precisely, tag the flavor of the D mesons.\par 

\vskip 3 truemm

- As for the decays $\overline{D^0}(D^0) \to f_i$, they can be determined from the $D^{*+}\to D^0\pi^+$ and $D^{*-}\to \overline{D^0}\pi^-$ decays, where the charge of the pion identifies the flavor of the $D^0$ meson. The $D^0-\overline{D^0}$ mixing leads to a negligible effect in the extraction of the Branching Fractions, as expected in the SM.

\vskip 10pt
In summary, we are left with two equations and three unknowns. In principle,  $\Gamma[B^+\to \overline{D^0}K^+  \to f_i K^+]$ could be also obtained experimentally. The corresponding number of events is 
\be
\qquad N_{B^+\to \overline{D^0}K^+  \to f_i K^+} = N_{B^+} \times Br(B^+ \to \overline{D^0}K^+)\times Br(\overline{D^0} \to f_i) \qquad
\label{eq:12}
\ee

\noindent Since the individual branching fractions are known, should one know precisely the initial number of $B^+$, one would extract the $N_{B^+\to \overline{D^0}K^+ \to f_i K^+} $. At the Z-pole $N_{B^+}$ is known with a precision of $\sim2\%$. Although this is not too bad and might be improved at FCCee, it is preferable to keep $ N_{B^+\to \overline{D^0}K^+  \to f_i K^+} $ as a free parameter, in which case one would need to use an additional final state $f_i$ as is discussed in \cite{ADS}.
\vskip 10pt
\noindent One also derives the CP asymetry :
\be
A_{CP} = \frac{|A|^2-|\overline{A}|^2}{|A|^2+|\overline{A}|^2} = \frac{- 2{\cal R}_{B\to f_iK}\sin\Delta_{f_i} \sin\Phi_{f_i}}{1+{\cal R}_{B\to f_iK}^2 +2{\cal R}_{B\to f_iK}\cos\Delta_{f_i} \cos\Phi_{f_i}} \qquad \qquad
\label{eq:13}
\ee

\noindent Let us now investigate further the three types of decays mentioned above.

\subsection{Using Doubly Cabibbo Suppressed $\overline{D^0}$ decays}

\vskip 2 truemm

In the following we consider the decay $B^+\rightarrow \overline{D^0}K^+ \rightarrow (K^-\pi^+)_{\overline{D^0}}K^+$ and 
$B^+\rightarrow D^0K^+ \rightarrow (K^-\pi^+)_{D^0}K^+$, where the former decay channel is Doubly Cabibbo Suppressed for the decay $\overline{D^0} \rightarrow K^-\pi^+$ and the latter Cabibbo Allowed for $D^0 \rightarrow K^-\pi^+$. Using this type of decays is known as the ADS method \cite{ADS}.
The complete amplitudes read :
$$A_{DCS} = a_{DCS}e^{i\alpha_1} e^{i\Phi_{DCS}} \ , \qquad A_{CA} = a_{CA}e^{i\alpha_2} e^{i\Phi_{CA}} \qquad \qquad$$
\be
\qquad \ \ \ \overline{A}_{DCS} = a_{DCS}e^{i\alpha_1} e^{-i\Phi_{DCS}}\ , \qquad \overline{A}_{CA} = a_{CA}e^{i\alpha_2} e^{-i\Phi_{CA}} \qquad \qquad \ \ \ 
\label{eq:14}
\ee

\vskip 2 truemm

\noindent where the subscripts DCS and CA stand for Doubly Cabibbo Suppressed and Cabibbo Allowed, respectively.\par
The CP violating phases in  (\ref{eq:14}) read,
\be
\qquad \Phi_{DCS} = \arg(V_{cb}^*V_{us}V_{us}^*V_{cd}) \ , \qquad \Phi_{CA} = \arg(V_{ub}^*V_{cs}V_{cs}^*V_{ud}) \qquad
\label{eq:15}
\ee

\vskip 2 truemm

\noindent and one gets
$$\Phi_{K^-\pi^+} = \Phi_{DCS} - \Phi_{CA} = \arg \left(V_{cb}^*V_{cd} \over V_{ub}^*V_{ud}\right) = - \arg \left(V_{ub}^*V_{ud} \over V_{cb}^*V_{cd}\right)$$
\be 
\qquad \qquad \qquad \qquad = \pi - \arg \left(- {V_{ub}^*V_{ud} \over V_{cb}^*V_{cd}}\right) = \pi - \gamma \qquad \qquad \qquad \qquad \ 
\label{eq:16}
\ee

\noindent where $\gamma$ is one of the angles of the usual Unitarity Triangle $UT_{db}$ of Fig. 1. Note also that $\Phi_{DCS}$ and $\Phi_{CA}$ are convention dependent, while $\Phi_{DCS} - \Phi_{CA}$ is not.\par

\vskip 2 truemm

One then writes the decay widths $\Gamma [B^+\rightarrow (K^-\pi^+)_{D}K^+]$ and $\Gamma [B^-\rightarrow (K^+\pi^-)_{D}K^-]$,
$$\Gamma [B^+\rightarrow (K^-\pi^+)_{D}K^+] = |A_{DCS} + A_{CA}|^2 \qquad \qquad \qquad \qquad $$
\be
\qquad \qquad \Gamma [B^-\rightarrow (K^+\pi^-)_{D}K^-] = |\overline{A}_{DCS} + \overline{A}_{CA}|^2 \qquad \qquad \qquad 
\label{eq:17}
\ee

\vskip 2 truemm

\noindent Thus, as discussed in the section above, with $\Delta = \alpha_1 - \alpha_2$, one has
$$\Gamma [B^+\rightarrow (K^-\pi^+)_{D}K^+]  =  a_{DCS}^2 + a_{CA}^2 +\ 2a_{DCS}a_{CA} [-\cos\Delta \cos\gamma - \sin\Delta \sin\gamma]$$
\be
 \Gamma [B^-\rightarrow (K^+\pi^-)_{D}K^-] = a_{DCS}^2 + a_{CA}^2 +\ 2a_{DCS}a_{CA} [-\cos\Delta \cos\gamma + \sin\Delta \sin\gamma] 
\label{eq:18}
\ee

\noindent where
$$a_{DCS}^2 = \Gamma(B^+ \to \overline{D^0}K^+)\times \Gamma(\overline{D^0} \to K^-\pi^+)
= \Gamma(B^- \to D^0K^-)\times \Gamma(D^0  \to K^+\pi^-)$$
\be
a_{CA}^2 = \Gamma(B^+ \to D^0 K^+)\times \Gamma(D^0  \to K^-\pi^+)
= \Gamma(B^- \to \overline{D^0} K^-)\times \Gamma(\overline{D^0}  \to K^+\pi^-) \ \ \ 
\label{eq:19}
\ee

\vskip 2 truemm

\noindent Thus, 
$$|A|^2 = \Gamma[B^+ \to (f_i)_D K^+] = \Gamma[B^+\to \overline{D^0}K^+ \to f_i K^+] \qquad \qquad$$  
$$ \times [ 1 + {\cal R}_{B\to f_iK}^2 + 2{\cal R}_{B\to f_iK} (- \cos\Delta_{f_i} \cos\gamma - \sin\Delta_{f_i} \sin\gamma )] \qquad \qquad$$ 
$$|\overline{A}|^2 = \Gamma[B^- \to (f_i)_D K^-] = \Gamma[B^+\to \overline{D^0}K^+  \to f_iK^+] \qquad \qquad$$ 
\be 
\qquad \times [1 + {\cal R}_{B\to f_iK}^2 + 2{\cal R}_{B\to f_iK}(- \cos\Delta_{f_i} \cos\gamma + \sin\Delta_{f_i} \sin\gamma )] \qquad \qquad \ 
\label{eq:20}
\ee

\vskip 3 truemm

\noindent where $f_i =K^-\pi^+$.\par 

We rewrite equations (\ref{eq:19}) in terms of ${\cal R}$ and ${\cal R}_{D\to f_i}$ as defined in equation  (\ref{eq:7}),
$$\Gamma [B^+\rightarrow (f_i)_{D}K^+] \ = \ \Gamma[B^+\to \overline{D^0}K^+  \to f_i K^+]$$  
$$\times [ 1 + ({\cal R}{\cal R}_{D\to f_i})^2 + 2{\cal R}{\cal R}_{D\to f_i} (- \cos\Delta_{f_i} \cos\gamma - \sin\Delta_{f_i} \sin\gamma )] \qquad \qquad$$  

$$\Gamma [B^-\rightarrow (\overline{f_i})_{D}K^-] \ = \ \Gamma[B^+\to \overline{D^0}K^+  \to f_iK^+]$$ 
\be 
\qquad \times [1 + ({\cal R}{\cal R}_{D\to f_i})^2 + 2{\cal R}{\cal R}_{D\to f_i}(- \cos\Delta_{f_i} \cos\gamma + \sin\Delta_{f_i} \sin\gamma )] \qquad 
\label{eq:21}
\ee

\vskip 3 truemm

\noindent As discussed in the previous section, ${\cal R}$ can be determined experimentally using the $B^\pm$ using semileptonics decays of the $\overline{D^0}(D^0)$ and ${\cal R}_{D\to f_i}$ can be determined from the $D^{*\pm}$ decays, where the charge of the pion identifies the flavor of the $D^0$ meson.

\vskip 10pt
Therefore $\gamma$ can be extracted from equation (\ref{eq:20}) provided the total number of produced $B^\pm$ is known in order to evaluate $\Gamma[B^+\to \overline{D^0}K^+  \to f_i K^+]$ following equation (\ref{eq:12}).\par 

\vskip 3 truemm

Should one leave $\Gamma[B^+\to \overline{D^0}K^+ \to f_i K^+]$ as a free parameter, we would be left with three unknowns for two equations and thus one should use additional modes. For example, one can consider the final state $\overline{f_i} = K^+\pi^-$. Using equations (\ref{eq:21}) and the relations in equation (\ref{eq:8}), one derives
$$\Gamma [B^+\rightarrow (\overline{f_i})_{D}K^+] \ = \ \Gamma[B^+\to \overline{D^0}K^+ \to f_i K^+]\ \qquad \qquad $$  
$$\times [ {\cal R}_{D\to f_i}^2 + {\cal R}^2 + 2{\cal R}{\cal R}_{D\to f_i} (- \cos{\overline \Delta}_{{\overline f}_i} \cos\gamma - \sin{\overline \Delta}_{{\overline f}_i} \sin\gamma )] \qquad \qquad$$   
$$\Gamma [B^-\rightarrow (f_i)_{D}K^-] \ = \ \Gamma[B^+\to \overline{D^0}K^+ \to f_iK^+]\ \qquad \qquad$$
\be  
\qquad\times [{\cal R}_{D\to f_i}^2 + {\cal R}^2 + 2{\cal R}{\cal R}_{D\to f_i}(- \cos{\overline \Delta}_{{\overline f}_i} \cos\gamma + \sin{\overline \Delta}_{{\overline f}_i} \sin\gamma )] \qquad \qquad
\label{eq:22}
\ee

\vskip 3 truemm

Compared to equation  (\ref{eq:20}), an additional parameter, ${\overline \Delta}_{\overline{f_i}}$, has been introduced. It concerns a strong phase difference as is discussed here below. Altogether, with equations (\ref{eq:21}) and (\ref{eq:22}), we have now four parameters, $\Gamma[B^+\to \overline{D^0}K^+  \to f_iK^+]$, $\Delta_{f_i}$, ${\overline \Delta}_{\overline{f_i}}$ and $\gamma$, and four equations. Thus $\gamma$ can be extracted.

\vskip 3 truemm

\vskip 10pt
\noindent For completeness, we write the CP asymetries
$${\cal A_{CP}}(f_i=K^-\pi^+) = \frac{\Gamma [B^+\rightarrow (K^-\pi^+)_{D}K^+] - \Gamma [B^-\rightarrow (K^+\pi^-)_{D}K^-]}
{\Gamma [B^+\rightarrow (K^-\pi^+)_{D}K^+] + \Gamma [B^-\rightarrow (K^+\pi^-)_{D}K^-]}$$
\be
{\cal A_{CP}}(f_i=K^-\pi^+) = \frac{- 2{\cal R}{\cal R}_{D\to f_i}\sin\Delta_{f_i} \sin\gamma}{1+{\cal R}^2{\cal R}_{D\to f_i}^2 - 2{\cal R}{\cal R}_{D\to f_i}\cos\Delta_{f_i} \cos\gamma} \qquad \qquad
\label{eq:23}
\ee

$${\cal A_{CP}}(\overline{f_i}=K^+\pi^-) = \frac{\Gamma [B^+\rightarrow (K^+\pi^-)_{D}K^+] - \Gamma [B^-\rightarrow (K^-\pi^+)_{D}K^-]}
{\Gamma [B^+\rightarrow (K^+\pi^-)_{D}K^+] + \Gamma [B^-\rightarrow (K^-\pi^+)_{D}K^-]}$$
\be
{\cal A_{CP}}(\overline{f_i}=K^+\pi^-) = \frac{- 2{\cal R}{\cal R}_{D\to f_i}\sin{\overline \Delta}_{\overline{f_i}} \sin\gamma}{{\cal R}_{D\to f_i}^2+{\cal R}^2 - 2{\cal R}{\cal R}_{D\to f_i}\cos{\overline \Delta}_{\overline{f_i}} \cos\gamma} \qquad \qquad
\label{eq:24}
\ee

\vskip 3 truemm

\noindent Comparing equations (\ref{eq:23}) and (\ref{eq:24}), the only parameters, which differ are $\Delta_{f_i}$ and ${\overline \Delta}_{\overline{f_i}}$.
$$\qquad  \Delta_{f_i} = \zeta_{\overline{D^0}} - \zeta_{D^0} + \delta_{1_{f_i}} - \delta_{2_{f_i}} = \Delta_\zeta + \Delta_{\delta_{f_i}} \qquad \qquad \qquad \qquad $$
\be
\qquad \qquad \ \ \ {\overline \Delta}_{\overline{f_i}} = \zeta_{\overline{D^0}} - \zeta_{D^0} + \delta_{1_{\overline{f_i}}} - \delta_{2_{\overline{f_i}}} = \Delta_\zeta + \Delta_{\delta_{\overline{f_i}}} \qquad \qquad \qquad
\label{eq:25}
\ee

\noindent where $\zeta_{\overline{D^0}}$ is related to $B^+\rightarrow \overline{D^0}K^+$ and $\zeta_{D^0}$ to $B^+\rightarrow D^0K^+$, while $\delta_{1_{f_i(\overline{f_i})}}$ is related to $\overline{D^0}\to f_i(\overline{f_i})$ and $\delta_{2_{f_i(\overline{f_i})}}$ to $D^0\to f_i(\overline{f_i})$.\par 

\vskip 3 truemm

It is important to underline that in general neither of these strong phase differences $\Delta_\zeta$ or  $\Delta_\delta$ are expected to vanish.\par 
As a final comment, the asymmetry ${\cal A_{CP}}(\overline{f_i}=K^+\pi^-)$ is expected to be very small since it suffers of a very low statistics for the mode $B^+\rightarrow D^0K^+ \rightarrow (K^+\pi^-)_{D^0}K^+$, as it has multiple suppression factors compared to $B^+\rightarrow \overline{D^0}K^+ \rightarrow (K^+\pi^-)_{\overline{D^0}}K^+$, namely $V_{ub}$ instead of  $V_{cb}$, as well as color and double Cabibbo suppressions. 

\vskip 10pt
In the following, we will not pursue the modes $\overline{D^0}(D^0) \rightarrow K^\pm\pi^\mp$. The reason is that they do not measure directly one of the phases of the ``flat'' triangle UT$_{sb}$ defined in (\ref{eq:1}), that is the main interest of the present paper.\par

\subsection{Using $\overline{D^0}(D^0)$ decays to CP eigenstates}

\subsubsection{$\overline{D^0}(D^0)$ decays to $\pi^+\pi^-$ }

\noindent Let us first examine the decay $\overline{D^0}(D^0) \rightarrow \pi^+\pi^-$, i.e. $CP = +$.
The complete amplitudes read 
$$ \qquad  A_1 = a_{\bar{D^0}(\pi\pi)}e^{i\alpha_{\bar{D^0}(\pi\pi)}} e^{i\Phi_{\bar{D^0}(\pi\pi)}} \ , \
A_2 = a_{D^0(\pi\pi)}e^{i\alpha_{D^0(\pi\pi)}} e^{i\Phi_{D^0(\pi\pi)}}  \qquad \qquad$$
\be
\qquad \overline{A_1} = a_{\bar{D^0}(\pi\pi)}e^{i\alpha_{\bar{D^0}(\pi\pi)}} e^{-i\Phi_{\bar{D^0}(\pi\pi)}} \ , \
\overline{A_2} = a_{D^0(\pi\pi)}e^{i\alpha_{D^0(\pi\pi)}} e^{-i\Phi_{D^0(\pi\pi)}} \ \ \ 
\label{eq:26}
\ee

\noindent where
\be
\qquad\Phi_{\bar{D^0}(\pi\pi)} = \arg(V_{cb}^*V_{us}V_{ud}^*V_{cd}) \  ,  \qquad       
\Phi_{D^0(\pi\pi)} = \arg(V_{ub}^*V_{cs}V_{cd}^*V_{ud})
\label{eq:27}
\ee

\noindent that gives
$$\Phi_{\bar{D^0}(\pi\pi)}-\Phi_{D^0(\pi\pi)} = \arg\left(\frac{V_{cb}^*V_{cd}}{V_{ub}^*V_{ud}}
\frac{V_{ud}^*V_{us}}{V_{cd}^*V_{cs}}\right)$$
$$= \arg\left(-\frac{V_{cb}^*V_{cd}}{V_{ub}^*V_{ud}}\right) +
\arg\left(- \frac{V_{ud}^*V_{us}}{V_{cd}^*V_{cs}}\right)$$
\be
\qquad \qquad \qquad \qquad \qquad \qquad = - \gamma + \gamma_{ds}  \simeq  -\gamma \qquad \qquad \qquad  \qquad \qquad \qquad \ 
\label{eq:28}
\ee

\noindent where $\gamma$ is one of the angles of the usual Unitarity Triangle $UT_{bd}$, and $\gamma_{ds}$ is a very small angle of the 3rd triangle $UT_{ds}$ in formulas (\ref{eq:1}) (\ref{eq:2}) and Fig. 1.\par 
One obtains the partial widths,

$$\Gamma [B^+\rightarrow (\pi\pi)_{D}K^+] = \Gamma[B^+\to \overline{D^0}K^+ \to (\pi\pi) K^+]$$  
$$\times [ 1 + {\cal R}^2 +2{\cal R} (\cos\Delta \cos\gamma + \sin\Delta \sin\gamma )]$$  

$$\Gamma [B^-\rightarrow (\pi\pi)_{D}K^-] = \Gamma[B^+\to \overline{D^0}K^+ \to (\pi\pi)K^+]$$  
\be
\qquad \qquad \qquad  \qquad \times [1 + {\cal R}^2 +2{\cal R}(\cos\Delta \cos\gamma - \sin\Delta \sin\gamma )]  \qquad \qquad  \qquad
\label{eq:29}
\ee

\noindent The asymetry reads

$${\cal A}_{CP}^+(\pi\pi) = \frac{\Gamma [B^+\rightarrow (\pi^+\pi^-)_{D}K^+] - \Gamma [B^-\rightarrow (\pi^+\pi^-)_{D}K^-]}
{\Gamma [B^+\rightarrow (\pi^+\pi^-)_{D}K^+] + \Gamma [B^-\rightarrow (\pi^+\pi^-)_{D}K^-]}$$
\be
\qquad \qquad \qquad{\cal A}_{CP}^+(\pi\pi) \simeq \frac{+ 2{\cal R}\sin\Delta \sin\gamma}{1+{\cal R}^2 +2{\cal R}\cos\Delta \cos\gamma}  \qquad \qquad \qquad \qquad
\label{eq:30}
\ee

\noindent where ${\cal R}$ is defined in equation (5) and $\Delta = \alpha_{\bar{D^0}(\pi\pi)}-\alpha_{D^0(\pi\pi)} \simeq\Delta_\zeta$. This is indeed so because  $\Delta_\delta \simeq 0$ in CP eigenstate decays of D mesons.

\subsubsection{$\overline{D^0}(D^0)$ decays to $K^+K^-$ }

We consider now the decays $B^+ \to D^0(\overline{D}^0) \to (K^+K^-)_{D} K^+$, i.e. $CP = +$. The complete amplitudes read :
$$A_1 = a_{\bar{D^0}(KK)}e^{i\alpha_{\bar{D^0}(KK)}} e^{i\Phi_{\bar{D^0}(KK)}} \        ,         \
A_2 = a_{D^0(KK)}e^{i\alpha_{D^0(KK)}} e^{i\Phi_{D^0(KK)}} \qquad$$
\be
\overline{A_1} = a_{\bar{D^0}(KK)}e^{i\alpha_{\bar{D^0}(KK)}} e^{-i\Phi_{\bar{D^0}(KK)}} \        ,         \
\overline{A_2} = a_{D^0(KK)}e^{i\alpha_{D^0(KK)}} e^{-i\Phi_{D^0(KK)}} 
\label{eq:31}
\ee

\vskip 3truemm

\noindent where the CP phases read
\be
\qquad \qquad \qquad \Phi_{\bar{D^0}(KK)} = \arg(V_{cb}^*V_{cs}) \        ,         \
\Phi_{D^0(KK)} = \arg(V_{ub}^*V_{us}) \qquad \qquad
\label{eq:32}
\ee
\be
\qquad \qquad \Phi_{\bar{D^0}(KK)}-\Phi_{D^0(KK)} = \arg\left(\frac{V_{cb}^*V_{cs}}{V_{ub}^*V_{us}} \right) = \pi + \gamma_{s} \qquad \qquad \qquad 
\label{eq:33}
\ee

\vskip 1truemm

\noindent The angle $\gamma_s$ is one of the angles of the Unitarity Triangle $UT_{sb}$ in Fig. 1. 

\vskip 3truemm

One obtains the partial widths from equation (\ref{eq:21}),

$$\Gamma [B^+\rightarrow (K^+K^-)_{D}K^+] = \Gamma[B^+\to \overline{D^0}K^+  \to (K^+K^-) K^+]$$  
$$\times [1 + {\cal R}^2 +2{\cal R} (- \cos\Delta \cos\gamma_s + \sin\Delta \sin\gamma_s )]$$  
$$\Gamma [B^-\rightarrow (K^+K^-)_{D}K^-] = \Gamma[B^+\to \overline{D^0}K^+  \to (K^+K^-)K^+]$$ 
\be 
\qquad \qquad \qquad \times [1 + {\cal R}^2 +2{\cal R}(- \cos\Delta \cos\gamma_s - \sin\Delta \sin\gamma_s )] \qquad \qquad  \qquad
\label{eq:34}
\ee

\noindent The asymetry reads
$${\cal A}_{CP}^+(K^+K^-) = \frac{\Gamma [B^+\rightarrow (K^+K^-)_{D}K^+] - \Gamma [B^-\rightarrow (K^+K^-)_{D}K^-]}
{\Gamma [B^+\rightarrow (K^+K^-)_{D}K^+] + \Gamma [B^-\rightarrow (K^+K^-)_{D}K^-]}$$
\be
\qquad \qquad \qquad {\cal A}_{CP}^+(K^+K^-) = \frac{+ 2{\cal R}\sin\Delta \sin\gamma_s}{1+{\cal R}^2 - 2{\cal R}\cos\Delta \cos\gamma_s} \qquad \qquad  \qquad \ \ 
\label{eq:35}
\ee

\vskip 3truemm

\noindent where ${\cal R}$ is defined in equation (7) and $\Delta = \alpha_{\bar{D^0}(KK)}-\alpha_{D^0(KK)}\simeq\Delta_\zeta$. 

\vskip 10pt
\noindent A similar result is obtained with the decay $\overline{D^0}(D^0) \to K_s\pi^0$. However due to the different CP sign ($\eta_{CP}=-1 $) for the eigenstate $K_s\pi^0$, the CP asymetry reads 
\be
\qquad \qquad \qquad  \qquad {\cal A}_{CP}^-(K_s\pi^0) = \frac{- 2{\cal R}\sin\Delta \sin\gamma_s}{1+{\cal R}^2 + 2{\cal R}\cos\Delta \cos\gamma_s} \qquad \qquad \ \ \ \ \ 
\label{eq:36}
\ee

\noindent with $\Delta_{K_s\pi^0} = \alpha_{\bar{D^0}(K_s\pi^0)}-\alpha_{D^0(K_s\pi^0)} \simeq \Delta_\zeta$.\par 
Using Eqns. (\ref{eq:35}) and (\ref{eq:36}) one can extract $\Delta(=\Delta_\zeta)$ and $\gamma_s$. This method had been proposed by Gronau, London and Wyler \cite{GLW} for the determination of the angle $\gamma$, since $\gamma_s \simeq - \gamma$, up to a very small correction. However, we are now interested in the angle $\gamma_s$ of the $UT_{sb}$ unitarity triangle.
Indeed, one gets 

$$\sin\Delta \sin\gamma_s = \frac{ {\cal A}_{CP}^-(K_s\pi^0)\times {\cal A}_{CP}^+(K^+K^-)}{{\cal A}_{CP}^-(K_s\pi^0)-{\cal A}_{CP}^+(K^+K^-)} \times \frac{1+{\cal{R}}^2}{\cal{R}}$$ 
\be
\qquad \ \ \ \cos\Delta \cos\gamma_s = - \frac{{\cal A}_{CP}^-(K_s\pi^0)+{\cal A}_{CP}^+(K^+K^-)}{{\cal A}_{CP}^-(K_s\pi^0)-{\cal A}_{CP}^+(K^+K^-)} \times \frac{1+{\cal{R}}^2}{2\cal{R}}\qquad \qquad 
\label{eq:37}
\ee

\vskip 3truemm

Thus $\gamma_s$ can be measured, though in general with a 8-fold ambiguity. However depending upon the values of $\cal{A}_{CP}^+$ and $\cal{A}_{CP}^-$ some solutions may not be physical and thus the ambiguities reduced. Several other $CP = -$ modes can be used to increase statistics, such as $\overline{D^0}(D^0) \to K_s\eta (\eta^\prime$) or $K_s\omega$.

\subsection{Convention of the difference of strong phases}

Here we comment on our convention on the difference of the strong phases $\Delta_{f_i}$, that differs from the convention in other works. Let us just consider the particular case of $B^\pm \to D^0(\overline{D}^0) K^\pm$, as made very explicit by Gronau \cite{GRONAU}.  

Let us compare the notation of our amplitudes  
\be
A[B^+ \to \overline{D}^0 K^+] = a_1 e^{i \alpha_1} e^{i \Phi_1} \ , \qquad A[B^+ \to D^0 K^+] = a_2 e^{i \alpha_2} e^{i \Phi_2} \ \ \qquad
\label{eq:38}
\ee 

\noindent and the CP-conjugated ones,
\be
A[B^- \to D^0 K^-] = a_1 e^{i \alpha_1} e^{- i \Phi_1} \ , \qquad A[B^- \to \overline{D}^0 K^-] = a_2 e^{i \alpha_2} e^{- i \Phi_2} \ \ \qquad
\label{eq:39}
\ee 

\vskip 2truemm

\noindent with Gronau convention, his formula (4),
\be
\ \ \ \ A[B^- \to D^0 K^-] =\ \mid A \mid \ , \qquad A[B^- \to \overline{D}^0 K^-] =\ \mid \overline{A}\mid e^{i \delta}\ e^{- i \gamma} \ \ \ \ \qquad
\label{eq:40}
\ee 

\noindent and its CP-conjugated,
\be
\ \ \ \ A[B^+ \to \overline{D}^0 K^+] =\ \mid A \mid \ , \qquad A[B^+ \to D^0 K^+] =\ \mid \overline{A}\mid e^{i \delta}\ e^{i \gamma} \ \ \ \ \qquad
\label{eq:41}
\ee 

\vskip 2truemm 

\noindent \noindent One has
\be
\qquad \qquad \qquad \delta = \alpha_2 - \alpha_1 = - \Delta \ , \qquad \qquad \Phi_2 - \Phi_1 = - \Phi = \gamma \qquad \qquad
\label{eq:42}
\ee

\noindent and finally
\be
\qquad \qquad \qquad \qquad \ \ \  \Delta = -\delta \ , \qquad \qquad \Phi = \Phi_1 - \Phi_2 = - \gamma \qquad \qquad \qquad
\label{eq:43}
\ee

\noindent So our FSI phase has the opposite sign to Gronau convention, and the CKM phase is the same in both conventions. The convention of Gronau is the same as the one used by HFAG \cite{HFAG}, where the phase $\delta$ is denoted by $\delta_B$. The opposite sign also holds for our general strong phase difference $\Delta_{f_i}$ and the one of HFAG $\delta_B + \delta_D$.

\section{Estimating the size of the modulus of amplitudes}

The ratio of amplitudes can be extracted experimentaly from the measured branching fraction (see Table 1).

$$ \begin{tabular}{lcccc}
\hline
 Mode & Br &  $\eta_{CP} $ & ${\cal R}$ & ${\cal R}_{D\to f_i}$ \\
\hline\hline
$B^+ \to \overline{D^0}K^+$ & $(3.63\pm 0.12)\cdot 10^{-4}$ &   & & - \\
$B^+ \to D^0K^+$ & $(3.57\pm 0.35)\cdot 10^{-6}$ &   &  0.099 $\pm$ 0.005 & -\\
$B^+ \to \overline{D^0}K^{*+}$ & $(5.3\pm 0.4)\cdot 10^{-4}$ &  & &  -\\
$B^+ \to D^0K^{*+}$ &$(3.1\pm 1.6)\cdot 10^{-6}$ &   & 0.076 $\pm$ 0.020& -\\
\hline
$\overline{D^0} \to K^+\pi^-$ & $(3.950\pm 0.031)\cdot 10^{-2}$ &   &  - & 0.0616$\pm$0.0015\\
$\overline{D^0} \to K^-\pi^+$ & $(1.50\pm 0.07)\cdot 10^{-4}$ &   & - & 1$6.23\pm 0.38$\\
$\overline{D^0} \to \pi^+\pi^-$ & $(1.455\pm 0.024)\cdot 10^{-3}$ & $+1$ & - & 1 \\
$\overline{D^0} \to K^+K^-$ & $(4.08\pm 0.06)\cdot 10^{-3}$ & $+1$ & - & 1\\
$\overline{D^0} \to K_s\pi^0$ & $(1.240\pm 0.022)\cdot 10^{-2}$ & $-1$ & - & 1 \\
$\overline{D^0} \to K_s\eta$ & $(5.090\pm 0.013)\cdot 10^{-3}$ & $-1$ & - & 1\\
$\overline{D^0} \to K_s\eta^\prime$ & $(9.49\pm 0.32)\cdot 10^{-3}$ & $-1$ & - & 1\\
$\overline{D^0} \to K_s\omega$ & $(1.11\pm 0.06)\cdot 10^{-2}$ & $-1$ & - & 1\\

\hline 

\end{tabular} $$

Table 1. The branching fraction of $B^+$ decays to final states $\overline{D^0}(D^0) K^+(K^{*+})$ and the subsequent $\overline{D^0}(D^0)$ decays \cite{PDG_2020}.

\vskip 10pt
\noindent Using the measured Branching Fractions \cite{PDG_2020}, one gets :
\be
\qquad \qquad \qquad {\cal R} = \sqrt { \frac{Br(B^+\rightarrow D^0K^+)}{Br(B^+\rightarrow \overline{D^0}K^+)} } = 0.099\pm 0.005 \qquad \qquad \qquad \ \
\label{eq:44}
\ee

\be
\qquad \qquad {\cal R}_{D\to K^-\pi^+} = \sqrt{\frac{Br(D^0\rightarrow K^-\pi^+)}{Br(\overline{D^0}\rightarrow K^-\pi^+)}} = 16.23\pm 0.38 \qquad \qquad \qquad
\label{eq:45}
\ee
$${\cal R}_{B\to (K^-\pi^+)K} = {\cal R}\times {\cal R}_{D\to K^-\pi^+} = 1.609\pm 0.092 \qquad \qquad $$

\noindent For any $CP$-eigentate, such as $f_{CP}=\pi^+\pi^-$, $K^+K^-$ or $K_s\pi^0$, ${\cal R}_{B\to (f_{CP})_DK} = {\cal R}$. Indeed, in the  former equation, $\frac{Br[D^0\rightarrow CP_{eigenstate}]}{Br[\overline{D^0}\rightarrow CP_{eigenstate}]}=1$ is assumed, since $|D^0_{CP\pm}>=\frac{1}{\sqrt{2}}[|D^0>\pm|\overline{D^0}>]$.


\vskip 10pt
\noindent As already mentioned in section 2.1, the ratio $Br(B^+\rightarrow D^0K^+)/Br(B^+\rightarrow \overline{D^0}K^+)$ should be measured even much more precisely at FCCee from the decays $B^+\rightarrow D^0K^+\rightarrow \ell^+\nu K^-K^+$ and $B^+\rightarrow \overline{D^0}K^+\rightarrow \ell^-\bar{\nu}K^+K^+$. The respective branching fractions are about 1.26$\cdot 10^{-7}$ and 1.29$\cdot 10^{-5}$. With the statistics expected at FCCee, one gets $5\cdot 10^4$ and $5\cdot 10^6$ events, respectively. Despite background levels orders of magnitude larger, it should be possible to extract the signal rather cleanly thanks to very accurate vertexing and particle identification for the kaons. A precision on $\delta ({\cal R})/{\cal R}$ at the level of $5\cdot 10^{-3}$ seems attainable.
\vskip 10pt

\section{Expected Asymetry Sensitivities at FCCee}

In this section, we investigate the sensitivity on $\gamma_s$, which one may expect at FCCee. The expected number of events at summarized in Table 2.

\par
\vskip 10pt

$$ \begin{tabular}{cccc}
\hline
&  & $\displaystyle {\mathrm {E_{cm} = 91.2\ GeV \ and\  \int L = 150 ab^{-1}}}$   &  \\

$\displaystyle \matrix{\displaystyle {\mathrm {\sigma (e^+e^- \to Z )}}
\\\displaystyle {\mathrm {nb}} \\}$
& 
$\displaystyle \matrix{\displaystyle {\mathrm {number}}
\\\displaystyle {\mathrm {of \ Z} }\\}$
& 
$\displaystyle \matrix{\displaystyle {\mathrm {f(Z\to B^+)}}
\\\displaystyle {\mathrm{}} \\}$
&
$\displaystyle \matrix{\displaystyle {\mathrm {Number \ of}}
\\\displaystyle {\mathrm{produced \ B^+ }} \\}$\\ 
\hline \hline 

$\displaystyle \sim 42.9$ &
$\displaystyle {\mathrm {\sim 6.4\ 10^{12}}}$ &
$\displaystyle {\mathrm {\sim0.061}} $ &
$\displaystyle \sim 3.9\ 10^{11} $\\ 
\hline
& & & \\
$\displaystyle \matrix{\displaystyle {\mathrm {B^+\ decay}}
\\\displaystyle {\mathrm {Mode}} \\}$
& 
$\displaystyle \matrix{\displaystyle {\mathrm {Decay}}
\\\displaystyle {\mathrm {Mode} }\\}$
& 
$\displaystyle \matrix{\displaystyle {\mathrm {Final}}
\\\displaystyle {\mathrm{State}} \\}$
&
$\displaystyle \matrix{\displaystyle {\mathrm {Number \ of}}
\\\displaystyle {\mathrm{B^+ \ decays}} \\}$\\ 
\hline \hline
&  & $\displaystyle {\mathrm {\overline{D^0}/D^0 \to non CP \ eigenstates}}$   &  \\

$\displaystyle \overline{D^0}K^+$ & 
$\displaystyle {\mathrm {\overline{D^0} \to K^+\pi^-}}$ &
$\displaystyle {\mathrm {K^+\pi^-K^+}} $ &
$\displaystyle \sim 5.6\ 10^6 $\\ 
$\displaystyle \overline{D^0}K^+$ &
$\displaystyle {\mathrm {\overline{D^0} \to K^-\pi^+}}$ &
$\displaystyle {\mathrm {K^-\pi^+K^+}} $ &
$\displaystyle \sim 2.1\ 10^4 $\\ 
$\displaystyle D^0K^+$ &
$\displaystyle {\mathrm D^0 \to K^-\pi^+}$ &
$\displaystyle {\mathrm K^-\pi^+K^+}$ &
$\displaystyle \sim 5.5 \ 10^4 $\\ 
$\displaystyle D^0K^+$ &
$\displaystyle {\mathrm D^0 \to K^+\pi^-}$ &
$\displaystyle {\mathrm K^+\pi^-K^+}$ &
$\displaystyle \sim 2.1 \ 10^2 $\\ 

&  & $\displaystyle {\mathrm {\overline{D^0}/D^0  \to CP \ eigenstates}}$   &  \\

$\displaystyle \overline{D^0}K^+$ &
$\displaystyle {\mathrm  {\overline{D^0} \to \pi^+\pi^-}}$ &
$\displaystyle {\mathrm {\pi^+\pi^-K^+}}$ &
$\displaystyle \sim 2.1 \ 10^5 $\\ 
$\displaystyle D^0K^+$ &
$\displaystyle {\mathrm {D^0\to \pi^+\pi^-}}$ &
$\displaystyle {\mathrm {\pi^+\pi^-K^+}}$ &
$\displaystyle \sim 2.0 \ 10^3 $\\ 

$\displaystyle \overline{D^0}K^+$ &
$\displaystyle {\mathrm  {\overline{D^0} \to K^+K^-}}$ &
$\displaystyle {\mathrm {K^+K^-K^+}}$ &
$\displaystyle \sim 5.8 \ 10^5 $\\ 
$\displaystyle D^0K^+$ &
$\displaystyle {\mathrm {D^0\to K^+K^-}}$ &
$\displaystyle {\mathrm {K^+K^-K^+}}$ &
$\displaystyle \sim 5.7 \ 10^3 $\\ 
$\displaystyle \overline{D^0}K^+$ &
$\displaystyle {\mathrm  {\overline{D^0} \to K_s\pi^0}}$ &
$\displaystyle {\mathrm {K_s\pi^0K^+}}$ &
$\displaystyle \sim 1.2 \ 10^6 $\\ 
$\displaystyle D^0K^+$ &
$\displaystyle {\mathrm {D^0\to K_s\pi^0}}$ &
$\displaystyle {\mathrm {K_s\pi^0K^+}}$ &
$\displaystyle \sim 1.2 \ 10^4 $\\ 

$\displaystyle \overline{D^0}K^+$ &
$\displaystyle {\mathrm  {\overline{D^0} \to K_s\eta}}$ &
$\displaystyle {\mathrm {K_s\eta K^+}}$ &
$\displaystyle \sim 5.0 \ 10^5 $\\ 
$\displaystyle D^0K^+$ &
$\displaystyle {\mathrm {D^0\to K_s\eta}}$ &
$\displaystyle {\mathrm {K_s\eta K^+}}$ &
$\displaystyle \sim 4.9 \ 10^3 $\\ 

$\displaystyle \overline{D^0}K^+$ &
$\displaystyle {\mathrm  {\overline{D^0} \to K_s\eta^\prime}}$ &
$\displaystyle {\mathrm {K_s\eta^\prime K^+}}$ &
$\displaystyle \sim 9.3 \ 10^5 $\\ 
$\displaystyle D^0K^+$ &
$\displaystyle {\mathrm {D^0\to K_s\eta^\prime}}$ &
$\displaystyle {\mathrm {K_s\eta^\prime K^+}}$ &
$\displaystyle \sim 9.1 \ 10^3 $\\ 

$\displaystyle \overline{D^0}K^+$ &
$\displaystyle {\mathrm  {\overline{D^0} \to K_s\omega}}$ &
$\displaystyle {\mathrm {K_s\omega K^+}}$ &
$\displaystyle \sim 1.1 \ 10^6 $\\ 
$\displaystyle D^0K^+$ &
$\displaystyle {\mathrm {D^0\to K_s\omega}}$ &
$\displaystyle {\mathrm {K_s\omega K^+}}$ &
$\displaystyle \sim 1.1 \ 10^4 $\\ 
\hline

\end{tabular} $$

Table 2. The mean expected number of produced $B^+$ decays to specific decay modes at FCCee at a center of mass energy of 91 GeV over 5 years with 2 detectors. These numbers have to be multiplied by 2 when including $B^-$ decays. The branching fractions of the PDG \cite{PDG_2020} have been used (only the $K_s$ decay to $\pi^+\pi^-$ is included). Note that the number of events is indicative since the exact numbers depend on the value of the strong phase differences $(\Delta)$ and the CKM phases involved, as seen in eqn. (\ref{eq:4}).

\subsection{Generic Detector acceptance and resolutions}

\noindent We define below a generic detector, the resolutions of which are  parametrized as follow  :

$${\rm Acceptance :} |\cos \theta| < 0.95$$
$${\rm Track\ p_T\ resolution :} {\sigma (p_T) \over p_T^2} = 2. \times 10^{-5} \ \oplus \ {1.2 \times 10^{-3}\over p_T \sin \theta}$$
$${\rm Track\ \phi , \theta \ resolution :} \mathrm{\sigma (\phi , \theta) \ \mu rad } = 18 \ \oplus \ {1.5 \times 10^{3} \over p_T\sqrt[3]{\sin \theta} }$$
\be
\qquad \qquad \qquad  {\rm Vertex \ resolution :} \mathrm{\sigma (d_{Im}) \ \mu m} = 1.8 \ \oplus \ {5.4 \times 10^{1} \over p_T\sqrt{\sin \theta} } \qquad \qquad
\label{eq:46}
\ee
$${\rm Vertex\ resolution :} <\sigma (\mathrm{d_{Im}})> \mathrm{bachelor\ K \ in}\  D_s K$$
$$<\sigma (\rm{d_{Im}})> \simeq 10\ \mu m$$
$${\rm Calorimeter\ resolution:} {\sigma (E) \over E} = {3 \times 10^{-2} \over \sqrt{E}} \ \oplus \ 5 \times 10^{-3}$$

\noindent where $\theta, \phi $ are the track polar and azymutal angle respectively, $p_T$ (in GeV) the track transverse momentum, $E$ the $e^\pm\ ,\gamma$ energy and $\mathrm{d_{Im}}$ the impact parameter. Although the tracking resolutions can be obtained with a combination of silicon vertex detector and a gazeous central tracking system, the calorimeter resolution requires a crystal type detector.\par

\vskip 10pt
\noindent The decays $B^+\rightarrow \overline{D^0}(D^0)K^+ \rightarrow (K^+K^-)_{D}K^+$ and $B^+\rightarrow \overline{D^0}(D^0)K^+ \rightarrow (K_s\pi^0)_{D}K^+$ have then been simulated using the resolutions in Equations (\ref{eq:46}).\par

\vskip2truemm
 
\includegraphics[scale=0.5]{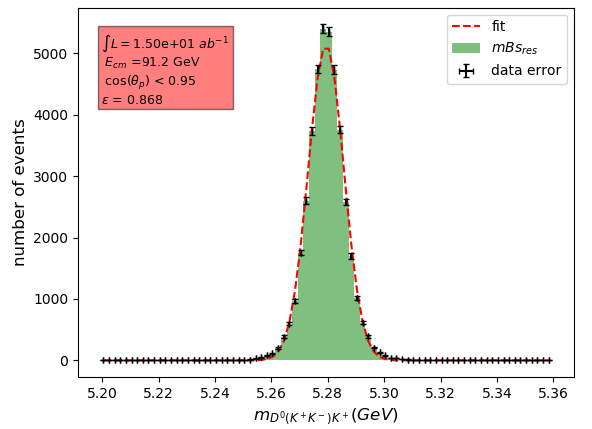}

\vskip3truemm

\begin{center}

Fig. 4. The $B^\pm$ resolutions  for the $D$ decays $K^+K^-$ in $e^+e^- \to Z \to B^+ \to D^0K^+ \to K^+K^-K^+$.

\end{center}

\includegraphics[scale=0.5]{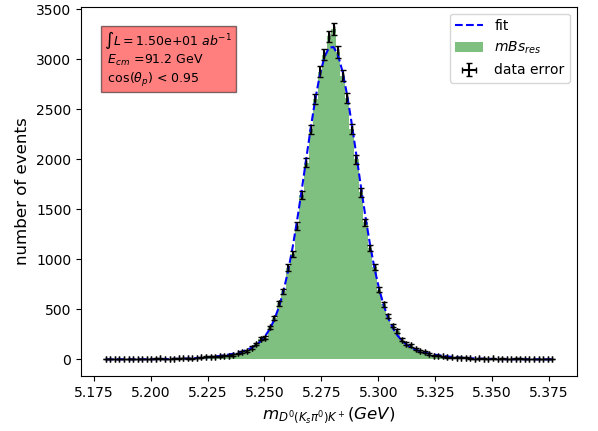}

\vskip3truemm

\begin{center}

Fig. 5. The $B^\pm$ resolutions  for the $D$ decays $K_s\pi^0$ in $e^+e^- \to Z \to B^+ \to D^0K^+ \to K_s \pi^0 K^+ \to \pi^+\pi^-\gamma \gamma K^+$.

\end{center}

\noindent As one can see from Figs. 4 and 5, the $B^\pm$ mass resolution is 6.4 MeV and 12.0 MeV for the $K^+K^-K^\pm$ and $K_s\pi^0K^\pm$ final states, respectively. For the latter decay, we use the constraint of the $D^0$ mass, which improves significantly the resolution. However, this resolution should worsen with a complete simulation, which will include accurately the flight distances of the resonances $B^\pm$, $D^0$ and $K_s$. It can also be noted that, using the values $\Delta = -130^\circ$ and $\gamma_s = 108^\circ$, one gets about the same number of events for both modes within $30\%$, once the acceptances are taken into account, in particular when requiring the $K_s$ to decay within a fiducial volume of $r = 1$ meter in the transverse direction and $z = \pm 1$ meter in the longitudinal direction. Such a fiducial volume for $K_s$ reconstruction implies that the tracking volume of the detector should be large enough, e.g. $r_{det.} > 2$ meter, $z_{det.} > \pm 2$ meter to enable efficient reconstruction of the $\pi^+ \pi^-$ pair. 
The tracking system should also consist of a large number of measurement layers, as shown in the appendix, which presents  an experimental study of the performance of $K_s$ reconstruction in a FCC-ee detector. As a final comment, one should note that in order to keep the background at a negligible level, an efficient particle identification system is required enabling to identify the kaons in a wide range from $\sim 1 - 30$ GeV.

\subsection{Sensitivity on the CKM angle $\gamma_s$}

To study the sensitivity, we assume a specific case with $\Delta = -130^\circ$ and $\gamma_s = 108^\circ$. With these figures, the asymmetries $A_{CP}^+(K^+K^-)$ and $A_{CP}^-(K_s \pi^0)$, as defined in equations (\ref{eq:35}) and (\ref{eq:36}) are -0.1489 and 0.1377, respectively.

\vskip2truemm

Using the expected number of events for the mode $B^\pm \to (K^+K^-)_D K^\pm$ and $B^\pm \to (K_s \pi^0)_D K^\pm$, and for a total integrated luminosity of $150\ ab^{-1}$ accumulated by the addition of all experiments at FCCee, the expected sensitivity for $\delta(\gamma_s)$ is about $2.7^\circ$ if $\Delta = -130^\circ$. Such a value for $\Delta$ corresponds to the strong phase in the decay $B^+ \to D^0 K^+$ as has been measured, and reported in \cite{PDG_2020}, $\delta_B = 130^\circ$, taking into account our convention $\Delta = -\delta_B$, as exposed in section 2.4.\par
One should note that the $\gamma_s$ sensitivity can vary according to the actual value of the strong angle difference $\Delta$, as can be seen in Table 3. The most favourable situation occurs when $\Delta = -90^\circ$, in which case $\delta (\gamma_s) \simeq 0.8^\circ$. When $\Delta = \pm \gamma_s$, the first equation in  (\ref{eq:37}) determines the sine-squared of this common (modulo the sign) angle, and is redundant with the second equation of (\ref{eq:37}), which determines the cosine-squared. Consequently, the equations (\ref{eq:37}) do not allow a simultaneous extraction of the angles $\Delta$ and $\gamma_s$. The same applies when $\Delta = 180^\circ \pm \gamma_s$. In the vicinity of these singularities, the experimental uncertainties on  ${\cal{A}}^+_{CP}( K^+ K^-)$ and on ${\cal{A}}^-_{CP} (K_s \pi^0)$ cause equation (\ref{eq:37}) to not always have real solutions -- for example, that would happen with a probability larger than $30 \%$ when the angles are within $8^\circ$ of a singularity. These cases are not considered in Table 3.\par

\vskip2truemm

These results can be improved further should one use additional $\overline{D^0}(D^0)$ CP eigenstates such as $K_s\eta$, $K_s\eta^\prime$, $K_s\omega\ldots$ and using the mode $B^\pm\rightarrow D_{CP}K^{*\pm}$ such that a sensitivity in the range 1$^\circ$-2$^\circ$ does not seem out of reach. Combining this with the possible direct measurements of $\alpha_s$ and $\beta_s$ with the modes $B_s\to D_s^\pm K^\mp$ and $B_s\to J/\psi \phi$, respectively, with the expected sensitivities \cite{ALEKSAN} $\delta(\alpha_s) \simeq 0.4^\circ$ and $\delta(\beta_s) \simeq 0.04^\circ$, one would directly measure the three angles of the Unitarity Triangle UT$_{sb}$ at FCCee, without assuming the unitarity of the CKM matrix.

\vskip3truemm

{\begin{center}
\begin{tabular}{|c|c|c|c|}
\hline 
$\Delta$ & ${\cal A}_{CP}^+(K^+K^-)$ & ${\cal A}_{CP}^-(K_s\pi^0)$ & $\delta(\gamma_s)$ \\ \hline \hline
$ -10^\circ$ & $-0.0306$ & $0.0345$ & $\sim 6.5^\circ$ \\ \hline
$ -30^\circ$ & $-0.0887$ & $0.0986$ & $\sim 2.8^\circ$ \\ \hline
$ -50^\circ$ & $-0.1377$ & $0.1489$ & $\sim 2.8^\circ$ \\ \hline  
$ -90^\circ$ & $-0.1868$ & $0.1868$ & $\sim 0.8^\circ$ \\ \hline   
$ -120^\circ$ & $-0.1668$ & $0.1570$ & $\sim 2.8^\circ$ \\ \hline  
$ -130^\circ$ & $-0.1489$ & $0.1377$ & $\sim 2.8^\circ$ \\ \hline
$ -150^\circ$ & $-0.0986$ & $0.0887$ & $\sim 2.8^\circ$ \\ \hline  
$ -170^\circ$ & $-0.0345$ & $0.0306$ & $\sim 6.5^\circ$ \\ \hline 
\end{tabular}
\end{center}}

Table 3. The experimental sensitivity for measuring the angle $\gamma_s$ at FCCee using the $B^\pm\rightarrow (K^+K^-)_{D}K^\pm$ and $B^\pm\rightarrow (K_s\pi^0)_{D}K^\pm$ decays for different values of the strong phase difference $\Delta$. The value $\gamma_s = 108^\circ$ has been used. Finally, one notes that should one use positive values for $\Delta$, all asymmetries will have opposite signs, but the sensitivities will remain the same.

\section{Conclusion}
\noindent We have investigated in this paper the sensitivity at FCCee of the angle $\gamma_s$ of the Unitarity Triangle UT$_{sb}$, defined in Fig. 3. This result is complementary to the ones obtained for the other angles of this triangle ($\alpha_s$ and $\beta_s$), as discussed in ref. \cite{ALEKSAN}. Altogether, FCCee should allow one to measure directly all the angles of the rather flat Unitarity Triangle UT$_{sb}$, constraining further the CP sector of the Standard Model.

{\section*{Appendix: reconstruction of $K_s \rightarrow \pi^+ \pi^-$ decays }}

The measurement of the decay $B^+ \rightarrow \overline{D^0} (D^0) K^+ \rightarrow ( K_s (\rightarrow \pi^+ \pi^-) \pi^0) K^+$ requires an efficient reconstruction of $K_s \rightarrow \pi^+ \pi^-$ decays, up to typically one meter from the interaction point (IP). 
This reconstruction, in a FCC-ee detector, has been studied using Monte-Carlo events that were passed through a fast simulation of the tracking system of the experiment based on DELPHES \cite{DELPHES}, and subsequently analysed within the FCCAnalyses framework \cite{HELSENS}. The PYTHIA8 Monte-Carlo generator was used to simulate the production of $b \bar{b}$ events at $\sqrt{s} = 91.2$~GeV, in which one $b$ quark hadronises into a charged $B$ meson that decays into the mode of interest, while the other $b-$leg fragments and decays inclusively. The decay chain of the $B^\pm$ mesons was performed with the EVTGEN program. The resulting charged particles were turned into simulated tracks using a fast tracking software implemented in DELPHES. It relies on a full description of the tracker geometry - the vertex detector and the drift chamber of the IDEA detector \cite{FCC}, which provide resolutions similar to the ones given in Section 4.1, being used here. The software accounts for the finite detector resolution and for the multiple scattering in each tracker layer and determines  the (non diagonal) covariance matrix of the helix parameters that describe the trajectory of each charged particle. This matrix is then used to produce a smeared 5-parameters track, for each charged particle emitted within the angular acceptance of the tracker. \\

The reconstruction of $K_s \rightarrow \pi^+ \pi^-$ heavily relies on the reconstruction of displaced  vertices. Here, a standalone code is used to fit a given set of tracks to a common vertex \cite{BEDESCHI}, under the assumption that the trajectories be perfect helices: the track parameters are varied according to their covariance matrix so that the tracks can meet at a common point, and the vertex coordinates $(x, y,z)$, together with their covariance matrix, are obtained from a $\chi^2$ minimisation. The vertexing program \cite{BEDESCHI} offers the possibility to ``guide" the first iteration of the fit by providing a guessed value of the radial position of the vertex, which can be inferred from the radial positions of the innermost hit of all the fitted tracks. This functionality is particularly useful when fitting tracks that made no hit in the ultra-precise silicon layers of the vertex detector surrounding the beam-pipe. \\

{\subsection*{Description of the algorithm}}

The first step in the $K_s$ reconstruction consists in finding the primary vertex and the primary tracks attached to this vertex (and consequently, the ``secondary" tracks which will be used to search for $K_s$ candidates). The identification of the primary vertex and of the primary tracks follows closely 
the LCFI+ algorithm \cite{SUEHARA}. All tracks that have been reconstructed are first fitted to a common vertex, using the knowledge of the beam-spot as an external constraint. The track which gives the largest contribution to the $\chi^2$ of the fit, if this contribution is larger than a given cut (here set to $25$), is then removed from the list of tracks, and the vertex fit is run again. This procedure is iterated until the contribution of all tracks to the $\chi^2$ is below the cut, providing the primary vertex and the ``primary" tracks attached to it. The tracks that are not attached to the primary vertex are labelled as ``secondary tracks". This selection of secondary tracks has a typical purity of more than $97 \%$, and is very efficient for tracks that come from a $K_s$ decay, in the decay chain considered here.

\begin{figure}[htb]
  \centering
  \begin{tabular}{cc}
  \includegraphics[width=0.48\columnwidth]{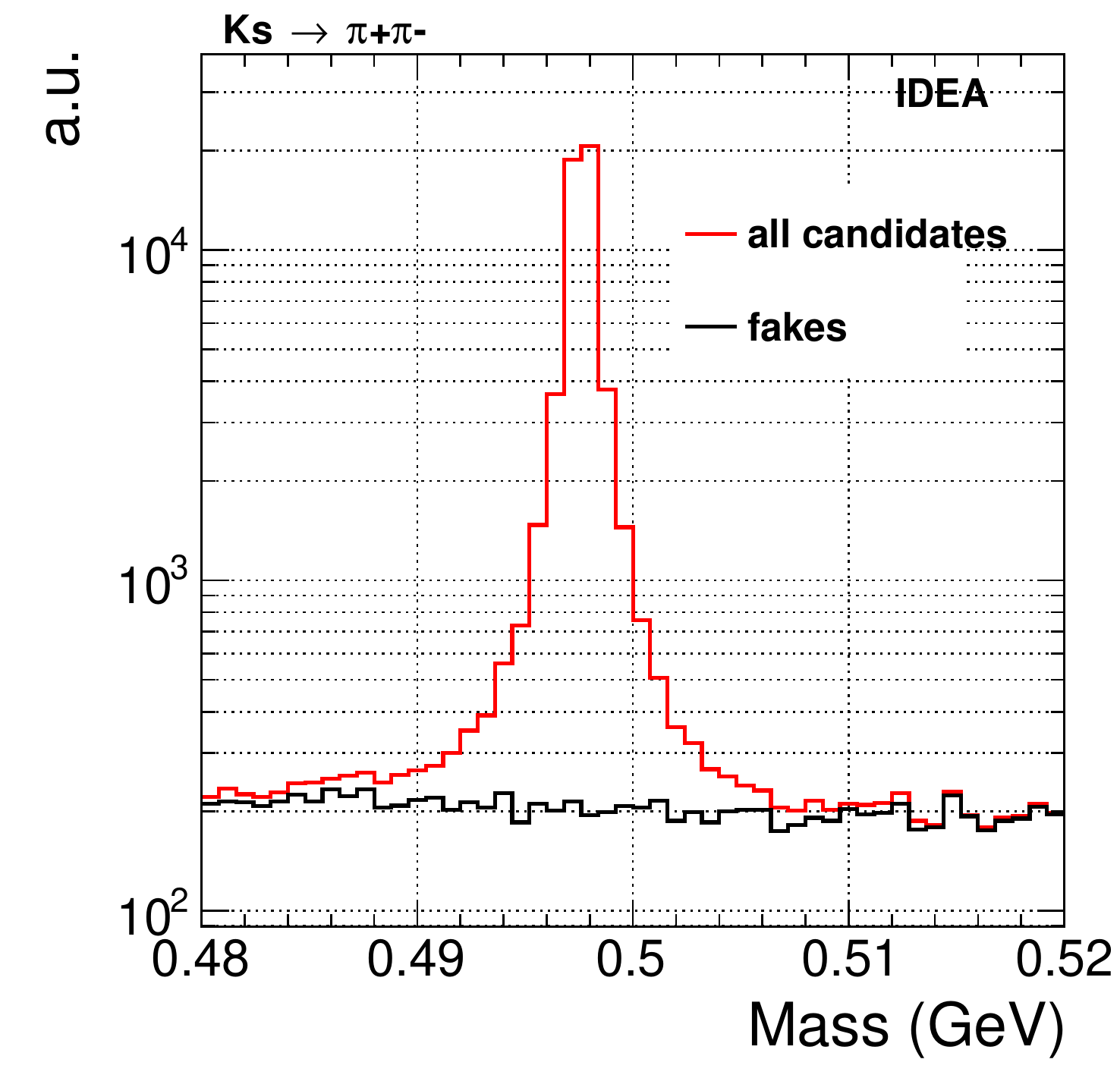}  &
  \includegraphics[width=0.48\columnwidth]{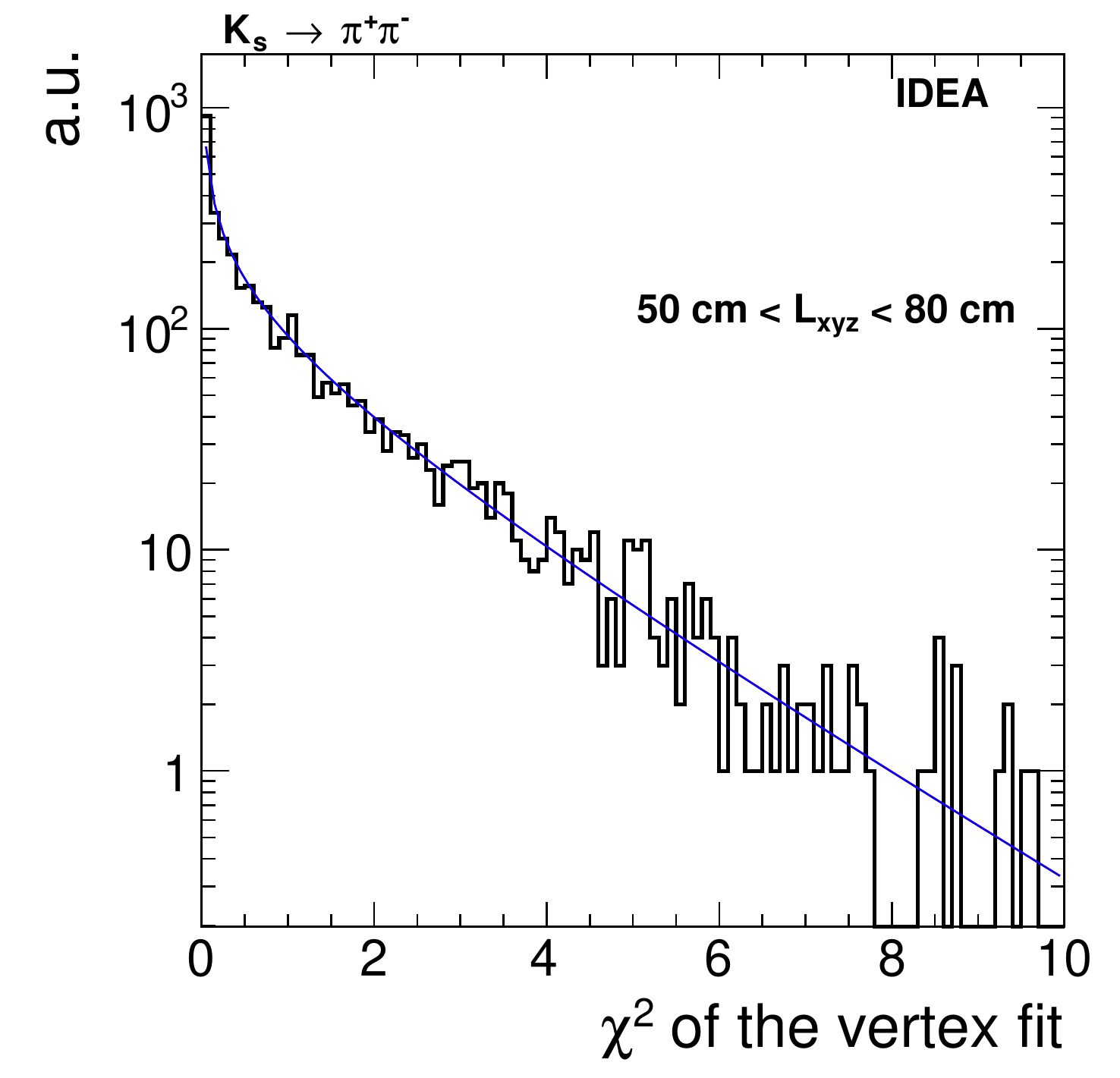} 
  \end{tabular}
\end{figure}
 Fig. 6. Distribution of the mass of reconstructed $K_s$ candidates in $Z \rightarrow b \bar{b}$ events which contain a $B^+ \rightarrow (K_s \pi^0)_D K^+$ decay, for all candidates (red histogram) and for fake candidates only (black histogram). Right: Distribution of the $\chi^2$ of the vertex fit of reconstructed candidates, that are matched to a genuine $K_s$ that decays at a distance between $50$~cm and $80$~cm from the interaction point. The overlaid curve represents a  $\chi^2$ function with one degree of freedom (the number of degrees of freedom of the two-tracks fit being equal to one.

\vskip 5 truemm

The identified secondary tracks are used in the second step of the algorithm. All combinations of two opposite charge tracks are built to form $K_s$ candidates. For each combination, the two tracks are fitted to a common vertex, and  the momenta of the tracks at this vertex are used to reconstruct the mass $M$ of the candidate, assigning the pion mass to each leg. Candidates are kept if they fulfil the loose mass cut $0.48 < M < 0.52$~GeV, and if the $\chi^2$ of the vertex fit is smaller than $10$. The latter requirement selects more than $99.4 \%$ of the $K_s$ decays in the process of interest. The association of the reconstructed tracks to the Monte-Carlo particles is then used to determine whether a candidate is matched to a $K_s \rightarrow \pi^+ \pi^-$ decay (any $K_s$ being considered here, not necessarily the one produced in the $B^+ \rightarrow DK^+$ decay). By definition, fake candidates are those which are not matched to such a decay. The mass distribution of all $K_s$ candidates reconstructed in the $B^+ \rightarrow DK^+$ sample, that satisfy the loose cuts listed above, is shown in Fig. 6 left. A clear peak is observed at the nominal $K_s$ mass used in the Monte-Carlo, on top of a flat background of fake candidates. A fit of the mass distribution of genuine $K_s$ candidates by a sum of two Gaussian functions leads to a mass resolution of about $400$~keV when the $K_s$ decays within $30$~cm from the interaction point, reaching $1$ to $1.2$~MeV for $K_s$ that decay at more than $1$~m from the IP. The distribution of the $\chi^2$ of the vertex fit for candidates that are matched to a $K_s  \rightarrow \pi^+ \pi^-$ decay occuring in the middle of the drift chamber is shown in Fig. 6 right. It can be seen that the vertex fit behaves well, even for far detached vertices.

{\subsection*{Kaon reconstruction efficiency}}

\begin{figure}[htb]
 \centering
  \begin{tabular}{cc}
  \includegraphics[width=0.48\columnwidth]{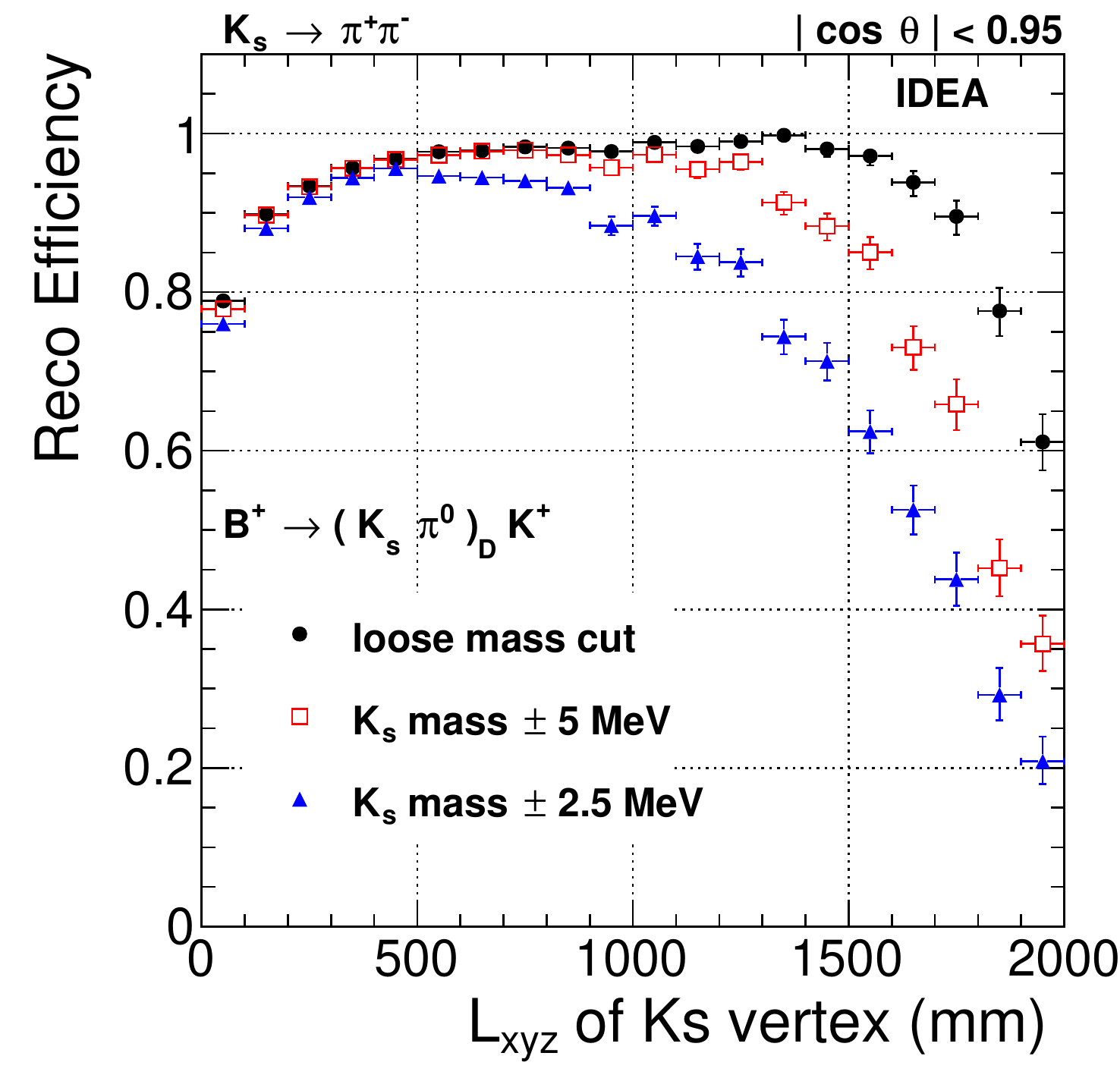} &
 \includegraphics[width=0.48\columnwidth]{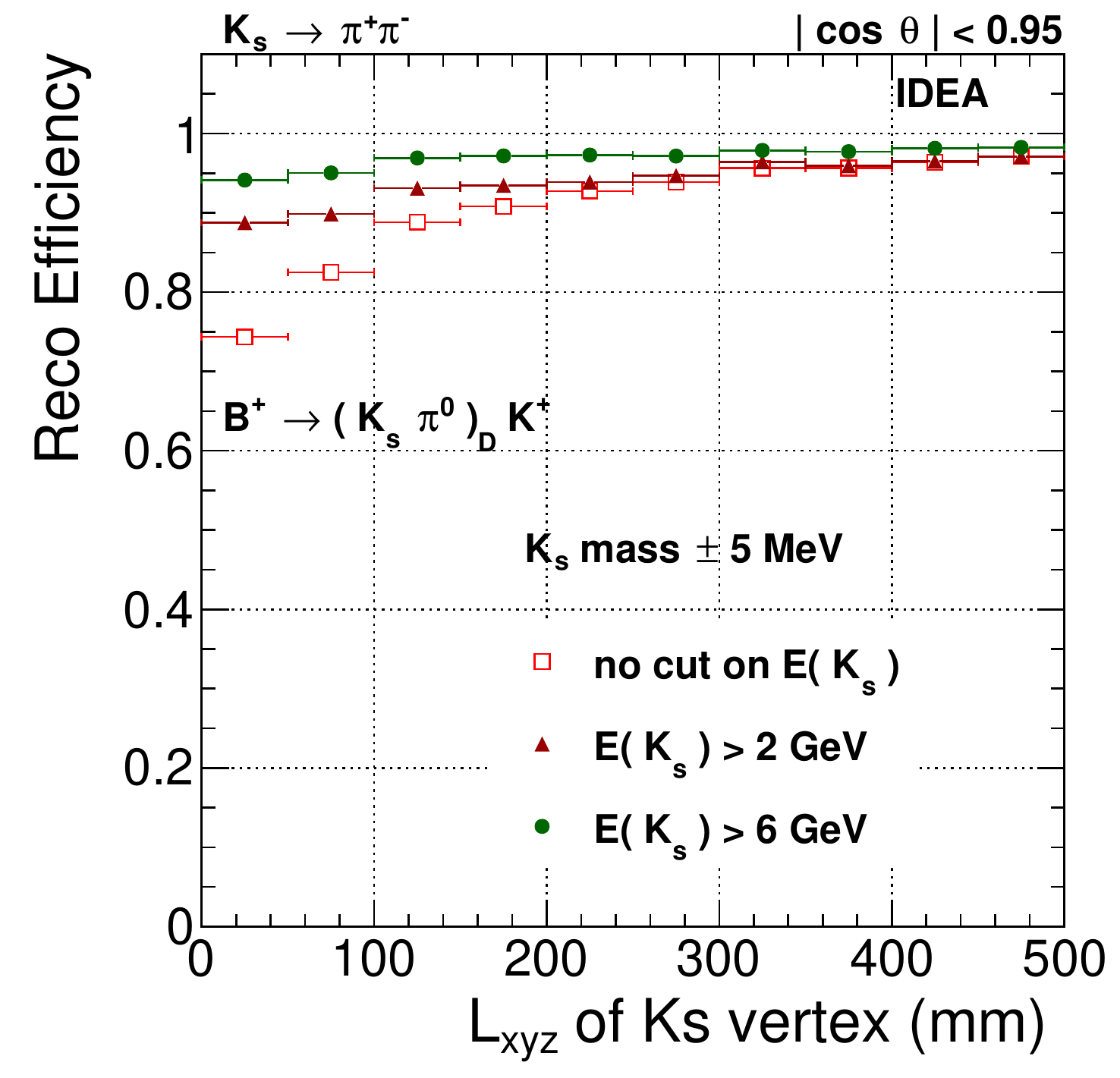} 
  \end{tabular}
\end{figure}
Fig. 7. Reconstruction efficiency of kaons from the $B^+ \rightarrow ( K_s \pi^0)_{D} K^+$ decay,
  as a function of the distance $L_{xyz}$ of the decay vertex of the $K_s$ to the IP, for (left) several cuts
  on the $K_s$ mass candidate, and (right) focusing on $L_{xyz} < 50$~cm and defining the efficiency with respect to kaons above a given energy.

\vskip 5 truemm

The reconstruction efficiency is shown in Fig. 7 left, as a function of the distance $L_{xyz}$ of the decay vertex of the $K_s$ to the interaction point. The efficiency is defined with respect to the $K_s$ that come from the  $B^+ \rightarrow ( K_s \pi^0)_{D} K^+$ decay, for which the Monte-Carlo daughter pions satisfy the acceptance cut $ | \cos \theta | < 0.95$. The efficiency obtained with a tighter mass cut of $\pm 5$~MeV or $\pm 2.5$~MeV around the nominal $K_s$ mass, which
is in line with the exquisite resolution mentioned previously, is also shown. The overall efficiency amounts to $90 \%$, $88.5 \%$ or $85 \%$ depending on the mass cut. With a cut of $\pm 5$~MeV, the efficiency remains above $80\%$ as long as the $K_s$ decays within $1.5$~m from the interaction point. At larger distances, the efficiency drops since the pion tracks only go through a small distance in the tracker. The drop in efficiency observed at small flight distances is due to the correlation of the flight distance with the $K_s$ momentum: the pion tracks from $K_s$ that decay close to the IP are softer and are more affected by multiple scattering or may lead to ``loopers". This correlation is illustrated in Fig. 7 right, which shows the efficiency to reconstruct kaons above a given energy. The small remaining inefficiency for energetic kaons that decay close to the interaction point is due to the occasional mis-assignment of the pion tracks to the primary vertex. \\

{\subsection*{Purity of the $K_s$ selection}}

The purity $P$ of the $K_s$ selection is studied in an inclusive sample of $Z$ bosons that decay hadronically. By definition, $1 - P$ corresponds to the fraction of reconstructed $K_s$ that are not matched to a Monte-Carlo $K_s \rightarrow \pi^+ \pi^-$. Figure 8 shows this fraction of mis-identified $K_s$, $1-P$, as a function of the distance $L_{xyz}$ of the decay vertex of the $K_s$ candidate to the interaction point. The purity is shown for two different mass cuts. Since the IDEA detector will provide a powerful identification of charged hadrons\footnote{This identification will be provided by measurements of the ionisation energy loss along the tracks, in the  drift chamber, and by the measurement of the time-of-flight of particles, for example in a dedicated layer outside the drift chamber.}, the purity is also shown under the assumption that charged pions can be perfectly separated from kaons and protons. At $L_{xyz} < 3$~cm, mis-identified $K_s$ usually correspond to random combinations of tracks coming from the decays of $\rho$, $\omega$, $D$ or other mesons (in $70\%$ of the cases, the two legs of the $K_s$ candidate do not come from the same parent). As expected, for candidates that decay within $1$~cm from the interaction point (first bin of Fig. 8 left), the purity is low, with a contamination of at least $25\%$. This contamination is drastically reduced as soon as the $K_s$ candidates decay at more than $1$ or $2$~cm from the IP, and this requirement can easily be applied in a $K_s$ selection as it is very efficient\footnote{For example, kaons from $B^+ \rightarrow ( K_s \pi^0)_{D} K^+$ in $Z \rightarrow b \bar{b}$ events decay at more than $1$~cm ($2$~cm) from the IP in more than $97\%$ ($94\%$) of the cases.}. Above $3$~cm, the fraction of mis-identified $K_s$ is at the per-mil level. In $85\%$ of the cases ($98\%$ at $L_{xyz} > 50$~cm), the two legs of the remaining fake candidates come from the same parent, and they usually correspond to $\Lambda$ baryons that decay into a proton and a pion. The $\Lambda$ lifetime being larger than that of the $K_s$, this explains the rise of the contamination with $L_{xyz}$, when  particle identification (PID) is not used in the $K_s$ identification, as shown in Fig. 8 right. However, this contamination is easily and efficiently suppressed by PID.

\begin{figure}[htb]
 \centering
  \begin{tabular}{cc}
  \includegraphics[width=0.48\columnwidth]{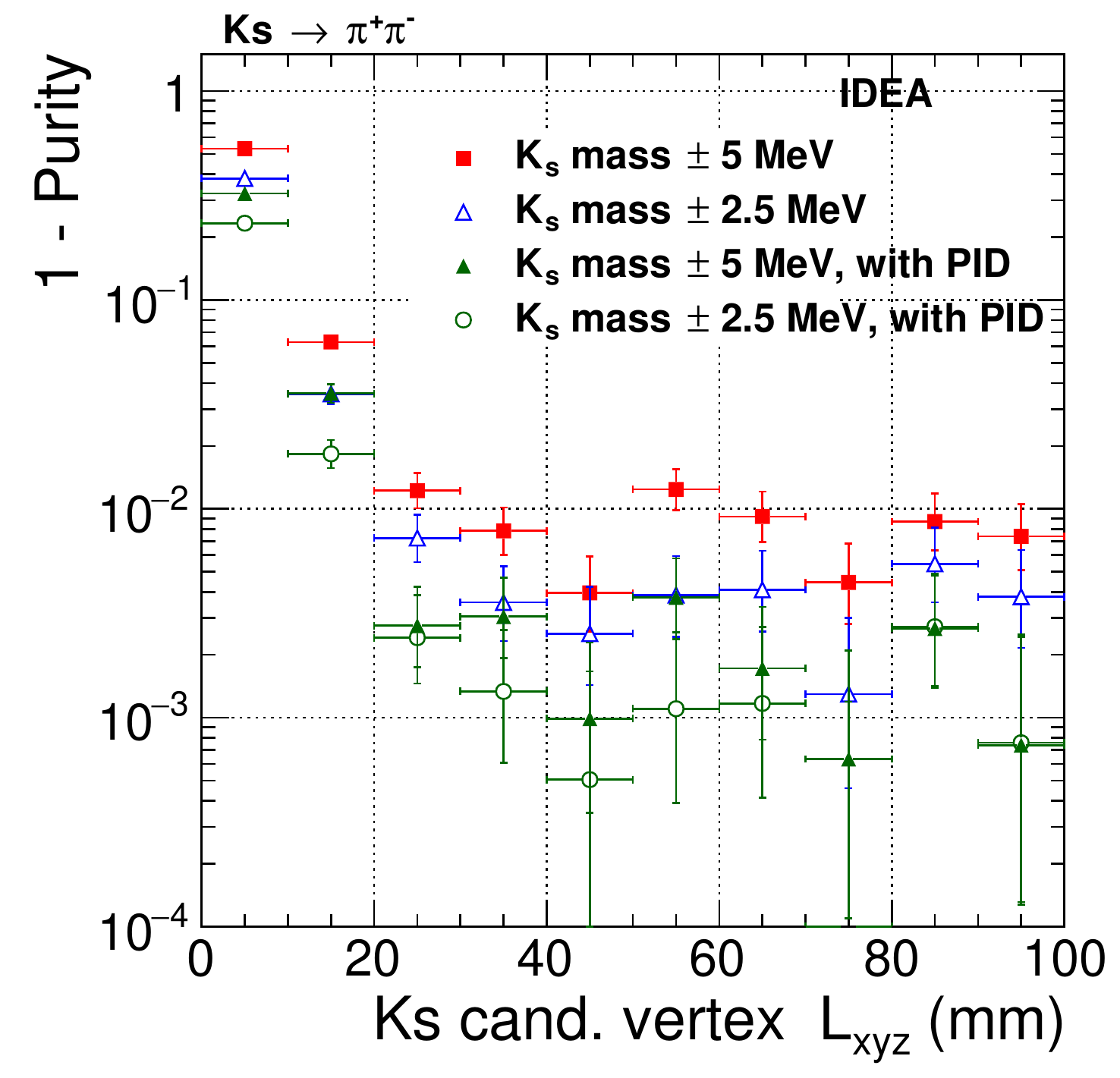} &
 \includegraphics[width=0.48\columnwidth]{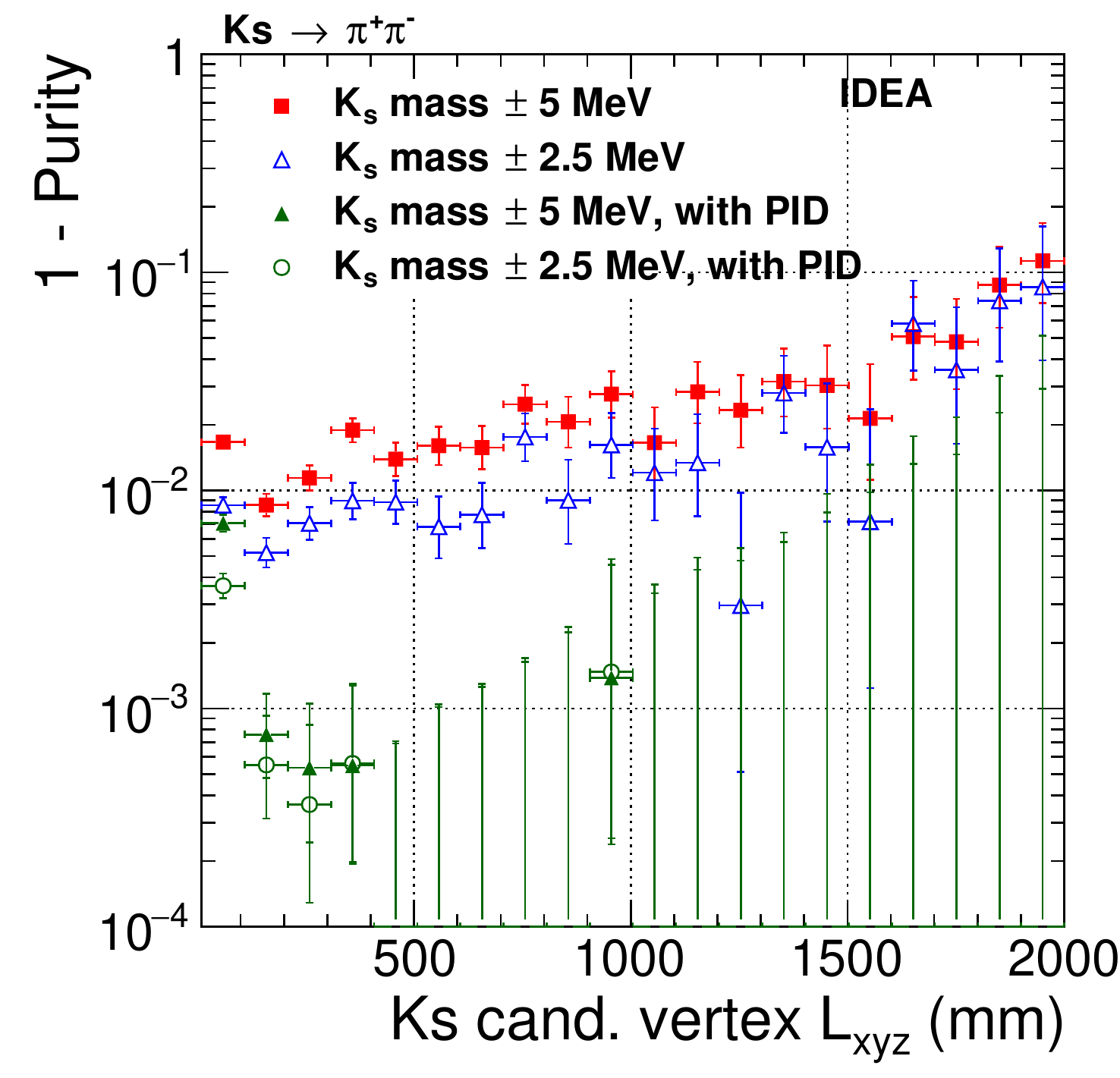} 
  \end{tabular}
\end{figure}
Fig. 8. Fraction of mis-identified $K_s$ in an inclusive $Z \rightarrow \mbox{hadrons}$ sample, as a function of the distance $L_{xyz}$ of the decay vertex of the candidate $K_s$ to the IP, for (left) $0< L_{xyz} < 10$~cm, and (right) for $1~\mbox{cm} < L_{xyz} < 2$~m. The different symbols correspond to different cuts used in the $K_s$ identification.

{\subsection*{Performance in a Full Silicon tracker}}

A ``continuous tracking", with a very large number of layers, as is the case in a drift chamber, is clearly a key for an efficient $K_s \rightarrow \pi^+ \pi^-$ reconstruction. To illustrate this point, the $B^+ \rightarrow ( K_s \pi^0)_{D} K^+$ events used above have been processed through a simulation of the CLD detector \cite{BACCHETTA}. The vertex detector of CLD is quite similar to the one of the IDEA detector, but its main tracker consists of a few layers of silicon sensors (for example six layers in the central region). Tracks made of at least five or four hits in the full tracker are used in the $K_s$ reconstruction algorithm described above. The resulting efficiencies are shown in Fig. 9. As expected, the performance dramatically worsens compared to the efficiency shown in Fig. 7 left. One clearly sees steps corresponding to the positions of the tracker layers. For example, as there are only four barrel layers at a radial distance larger than $40$~cm, the efficiency for selecting  5-hits tracks displaced by more than $40$~cm vanishes in the central region, explaining the step seen in Fig. 9 left. Some efficiency can be recovered by loosening the requirement on the minimum number of hits of the tracks to four, as shown in Fig. 9 right, at the expense of a likely increased rate of fake tracks\footnote{Fake tracks can not be studied with the fast simulation tool used here.}. Moreover, for $K_s$ that decay within  a few tens of centimeters from the IP, the efficiency is also lower than the one obtained with the IDEA drift chamber. This is because the material of the full silicon tracker is twice larger than the one of the IDEA tracker (drift chamber + vertex detector), leading to much larger effects from multiple scattering and worse resolutions on the $K_s$ vertex and on the $K_s$ mass. While some optimisations could be made to the CLD tracker, it is clear that an efficient reconstruction of $K_s \rightarrow \pi^+ \pi^-$ (and of long-lived particles in general) calls for a highly transparent tracker with  a very large number of layers. \\

\begin{figure}[htb]
 \centering
  \begin{tabular}{cc}
  \includegraphics[width=0.48\columnwidth]{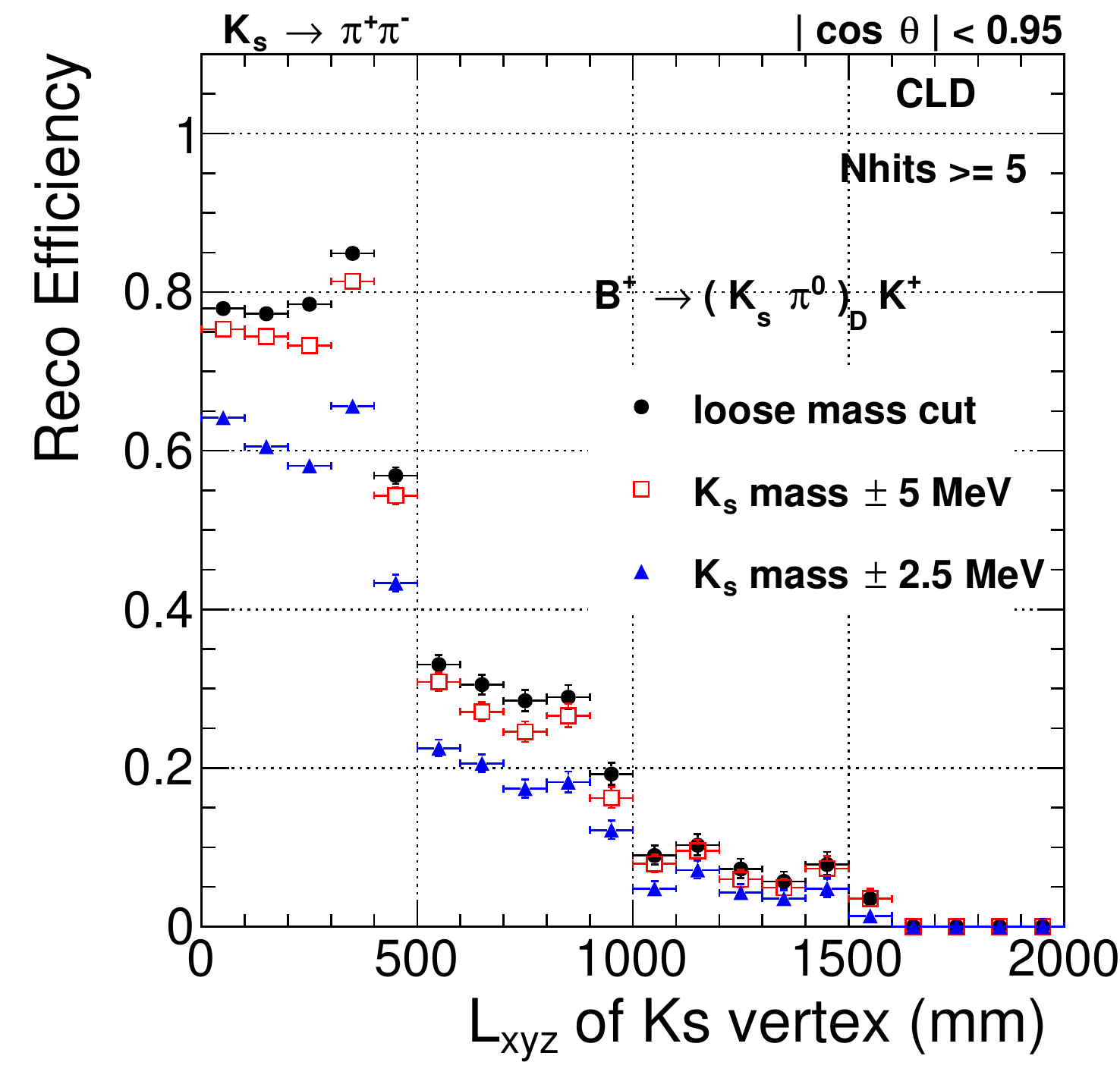} &
 \includegraphics[width=0.48\columnwidth]{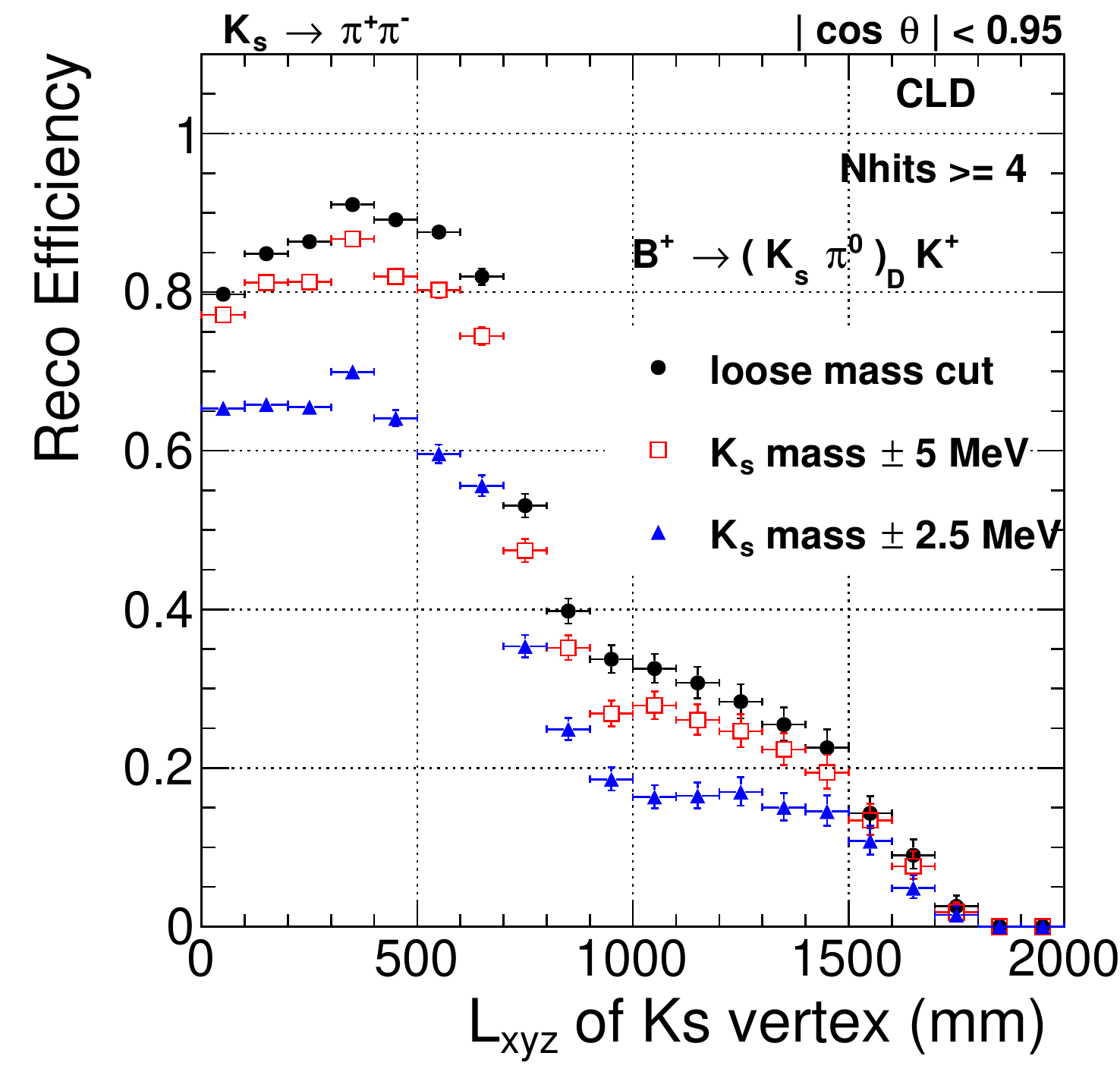} 
  \end{tabular}
\end{figure}
Fig. 9. Reconstruction efficiency of $K_s$ from the $B^+ \rightarrow ( K_s \pi^0)_{D} K^+$ decay,
  as a function of the distance to the IP of the decay vertex of the $K_s$, using tracks reconstructed in the CLD full silicon tracker, demanding at least five hits (left) or at least four hits (right) on the tracks.

{\subsection*{Acknowledgments}}
We wish to thank Franco Bedeschi for making his vertexing code available and for very useful discussions about the reconstruction of far displaced vertices. \\

\bibliographystyle{jhep}
\bibliography{biblio}

\begin{thebibliography}{99}

\bibitem{ALEKSAN} R. Aleksan, L. Oliver and E. Perez, arXiv:2107.02002 [hep-ph] (2021).
\bibitem{AKL} R. Aleksan, B. Kayser and D. London, Phys. Rev. Lett. {\bf 73} (1994) 18.
\bibitem{ADS} D. Atwood, I. Dunietz and A. Soni, Phys. Rev. Lett. {\bf 78} (1997) 3257.
\bibitem{PDG_2020} P.A. Zyla et al. [Particle Data Group], Prog. Theor. Exp. Phys. (2020) 083C01.
\bibitem{GLW} M. Gronau and D. London, Phys. Lett. B {\bf 253}, 483 (1991); M. Gronau and D. Wyler, Journal Phys. Lett. B {\bf 265}, 172 (1991).
\bibitem{GRONAU} M. Gronau, Phys. Lett. B {\bf 557} (2003) 198, arXiv:hep-ph/0211282.
\bibitem{HFAG} Heavy Flavor Averaging Group, Y. Amhis et al., arXiv:1909.12524 [hep-ex] (2019).
\bibitem{DELPHES} J. de Favereau et al., Journal of High Energy Physics {\bf 2014} (Feb. 2014).
\bibitem{HELSENS} C. Helsens and the FCC software group, https://github.com/HEP-FCC/FCCAnalyses.
\bibitem{FCC} FCC Collaboration, A. Abada et al., The European Physical Journal Special Topics {\bf 228} (2019) 261.
\bibitem{BEDESCHI} F. Bedeschi, Code available as part of the Delphes package, https://github.com/delphes/delphes.
\bibitem{SUEHARA} T. Suehara and T. Tanabe, Nucl. Instrum. Meth. A {\bf 808} (2016) 109.
\bibitem{BACCHETTA} N. Bacchetta et al., arXiv:1911.12230 (2019).

\end{thebibliography}

\end{document}